\documentclass[article, a4paper]{IEEEtran}

\ifCLASSINFOpdf
  \usepackage[pdftex]{graphicx}
\else
\fi

\usepackage{url}
\usepackage{breakurl}
\usepackage[cmex10]{amsmath}
\usepackage{amssymb}
\usepackage{units}
\usepackage{multirow}
\usepackage{rotating}
\usepackage{anyfontsize}
\usepackage{arydshln} 
\usepackage{textcomp}
\usepackage{color}
\usepackage{hyperref}

\usepackage{lipsum}
\usepackage[pscoord]{eso-pic}

\newcommand{\placetextbox}[3]{
  \setbox0=\hbox{#3}
  \AddToShipoutPictureFG*{
    \put(\LenToUnit{#1\paperwidth},\LenToUnit{#2\paperheight}){\vtop{{\null}\makebox[0pt][c]{#3}}}%
  }%
}%

\newcommand{\Link}{L}
\newcommand{\C}{C}

\newcommand{\Qi}{Q_i}

\newcommand{\Qni}{{\overline{Q}}_i}

\IEEEoverridecommandlockouts 
\begin{document}
\placetextbox{0.5}{1}{This is the author's version of an article that has been published in this journal.}
\placetextbox{0.5}{0.985}{Changes were made to this version by the publisher prior to publication.}
\placetextbox{0.5}{0.97}{The final version of record is available at \href{https://doi.org/10.1109/TWC.2018.2884914}{https://doi.org/10.1109/TWC.2018.2884914}}%
\placetextbox{0.5}{0.05}{Copyright (c) 2018 IEEE. Personal use is permitted.}
\placetextbox{0.5}{0.035}{For any other purposes, permission must be obtained from the IEEE by emailing pubs-permissions@ieee.org.}%

\title{Experimental Evaluation of Techniques to Lower Spectrum Consumption in Wi-Red\thanks{This work was partially supported by Regione Piemonte and the Ministry of Education, University, and Research of Italy in the POR FESR 2014/2020 framework, Call ``Piattaforma tecnologica Fabbrica Intelligente'', Project ``Human centered Manufacturing Systems'' (application number 312-36). The authors are with the National Research Council of Italy, Istituto di Elettronica e di Ingegneria dell'Informazione e delle Telecomunicazioni (CNR-IEIIT), I-10129 Torino, Italy (e-mail: \{name.surname\}@ieiit.cnr.it).}}

\author{Gianluca~Cena, \textit{Senior Member}, \textit{IEEE}, Stefano~Scanzio, \textit{Member}, \textit{IEEE}, and\\Adriano~Valenzano, \textit{Senior Member}, \textit{IEEE}
}

\maketitle

\begin{abstract}
Seamless redundancy layered atop Wi-Fi has been shown able to tangibly increase communication quality, hence offering industry-grade reliability.
However, it also implies much higher network traffic, which is often unbearable as the wireless spectrum is a shared and scarce resource.
To deal with this drawback the Wi-Red proposal includes suitable duplication avoidance mechanisms, which reduce spectrum consumption by preventing transmission on air of inessential frame duplicates.

In this paper, the ability of such mechanisms to save wireless bandwidth is experimentally evaluated.
To this purpose, specific post-analysis techniques have been defined, which permit to carry out such an assessment on a simple testbed that relies on plain redundancy and do not require any changes to the adapters' firmware.
As results show, spectrum consumption decreases noticeably without communication quality is impaired. 
Further saving can be obtained if a slight worsening is tolerated for latencies.
\end{abstract}


\section{Introduction}
Wireless communications are currently employed in many different contexts.
Besides personal connectivity and office/home environments, they are becoming increasingly popular also in other scenarios, like intelligent transportation systems \cite{IEEE-STD_1609-0_2014}, Internet of Things (IoT) \cite{RFC7554}, environmental monitoring \cite{2017_TWC_WSN}, and precision agriculture \cite{2015_IEEESensor_Agriculture}, to cite a few.
Although wireless networks have the potential to bring tangible benefits also to time-sensitive applications, they are quite unreliable and scarcely deterministic, and so are often deemed unsuitable in contexts like factory automation.
As a matter of fact, with the exception of WirelessHART and ISA100.11a \cite{Petersen2011}, real-time distributed control systems based on wireless technologies are currently quite unusual in industrial plants.

Unlike multimedia applications, where data can typically be buffered on both sides of a network connection, control applications (relying on, e.g., the request-response or producer-consumer data exchange paradigms \cite{2005_ProcIEEE}) demand that every single piece of information is delivered timely and reliably.
This can be hardly achieved with conventional wireless solutions, since they are typically based on \emph{random} access mechanisms, which unavoidably lead to collisions.
Behavior can be improved by using \emph{deterministic} access schemes, which prevent intra-network interference, e.g., Time Slotted Channel Hopping (TSCH) for IEEE 802.15.4 \cite{IEEE-STD_802-15-4_2015} and Hybrid-coordination-function Controlled Channel Access (HCCA) for IEEE 802.11 \cite{IEEE-STD_802-11_2016}, not counting the many proposals found in literature, for instance \cite{2007-TII-vas}\cite{2009_TWC_MACfree}\cite{2013_RTSS_RT-WiFi}\cite{TII16-EDF}.
It is worth noting that most of these mechanisms are unable  to face electromagnetic disturbance (including multipath fading effects) and, in several cases, not even interference of nearby non-compliant wireless devices.
These phenomena are unpredictable, and can be effectively counteracted by exploiting \emph{diversity}, in time, frequency, space, and so on.
For instance, automatic retransmission mechanisms are customarily included in any wireless protocol, frequency hopping is adopted by TSCH and Bluetooth, while multiple spatial streams are exploited in recent IEEE 802.11 versions.

In the following, we consider the use of IEEE 802.11, also known as \mbox{Wi-Fi}, for connecting devices in time-sensitive control applications.
The reasons of this choice are threefold: 
1) Wi-Fi is extremely popular, and consequently not expensive,
2) it features very high throughput, and 
3) it achieves complete interoperability with Ethernet.
We specifically focused on mechanisms based on time and frequency diversity (i.e., retransmissions and channel redundancy), both of which bring improvements on communication quality at the cost of higher spectrum consumption.
Techniques based on message scheduling, like HCCA or the like, were not taken into account as, to the best of our knowledge, there is currently little availability of commercial equipment offering proper support.

\emph{Seamless redundancy}, as defined by the Parallel Redundancy Protocol (PRP) \cite{2012-std-PRP}, is meant to increase availability of real-time Ethernet networks. 
Nonetheless, it can be also used to improve communication quality of wireless links.
In \cite{2012-WFCS-WoP1}, an arrangement was proposed, we denote PRP over Wi-Fi (PoW), which layers end-to-end seamless redundancy as per PRP atop IEEE 802.11, by using commercial \mbox{Wi-Fi} equipment, like access points (AP) and wireless adapters (WA), as well as specialized PRP devices (RedBoxes) \cite{redbox}.
Behavior, in terms of the fraction of frames delivered to destination correctly and timely, was experimentally shown to improve substantially.
Similar approaches were proposed for streaming phasor measurements in smart grids \cite{2016-SGC-Phasor}.

While relying on the same principles as PoW, the Wi-Fi Redundancy (Wi-Red) approach \cite{2016-tii-WiRed} is a link-level solution where PRP and the IEEE 802.11 Medium Access Control (MAC) mechanism are intertwined in order to improve performance further.
In particular, specific \emph{duplication avoidance} (DA) mechanisms are introduced to prevent transmission on air of identical copies of the same packet, when doing so is useless, so as to reduce network load.
Basically, Wi-Red can be seen as an holistic approach aimed at optimizing time and frequency diversity in Wi-Fi.
In theory, the same approach could be also used with wireless technologies other than Wi-Fi, like those employed in wireless sensor networks (WSN).
However, this is typically pointless, as WSNs are mainly aimed at ensuring low power consumption, and not high performance.
For example, solutions like TSCH offer time-frequency diversity through channel hopping, without the need to have a dual radio block, at the expense of larger transmission latencies.

This paper, which grounds on the preliminary work presented in \cite{2017-WFCS-Dupl}, experimentally evaluates the effects of DA mechanisms on the overall \emph{spectrum consumption} (combined traffic on all physical channels) of a redundant link between two wireless stations, and analyzes the trade-off between bandwidth saving and communication quality for a basic proactive heuristic.
With respect to \cite{2017-WFCS-Dupl}:
1) communication efficiency is evaluated considering the actual number of transmission attempts on air, inclusive of retries;
2) runs include more than twice samples and employ more realistic interfering traffic patterns;
3) assumptions on virtual analysis of proactive approaches are validated; and
4) both frame losses and more-than-duplex redundancy are considered.
The paper is organized as follows: Section~\ref{sec:redundancy} summarizes Wi-Fi seamless redundancy basics, while Section~\ref{sec:setup} describes our experimental testbed.
Quantitative evaluation of the ability of DA mechanisms to reduce spectrum consumption, based on measurements carried out on the testbed, is reported in Sections~\ref{sec:RDA} and \ref{sec:PDA}, which concern two types of approaches, that are reactive and proactive ones, respectively.
Finally, some conclusions are drawn in Section~\ref{sec:Conclusion}.

\section{Link-level Wi-Fi Seamless Redundancy}
\label{sec:redundancy}
In its simplest embodiment, Wi-Red corresponds to a thin layer, located just above Wi-Fi adapters and implemented by bringing software modifications to their device drivers.
However, to unleash all its potential, adapters' firmware has likely to be rewritten in part.

\subsection{Redundant Network Architecture}
An IEEE 802.11 Basic Service Set (BSS) is a group of wireless stations (STA) located in the same place, operating on the same channel, and enabled to exchange data among each other.
Similarly, a \emph{redundant BSS} (RBSS) is defined as a set of co-located \emph{redundant STAs} (RSTA) communicating on multiple physical channels according to seamless redundancy principles.
An RSTA comprises two or more sub-STAs, operating on distinct channels, and one \emph{link redundancy entity} (LRE), which coordinates their operations so as to achieve seamless redundancy.
Each sub-STA includes both a MAC and a radio block, and behaves mostly the same as a conventional STA. 
More details can be found in \cite{2016-tii-WiRed}.

Generally speaking, effectiveness of seamless redundancy in networks including multiple uncoordinated RSTAs is typically not as good as on point-to-point links.
In fact, intra-network interference they generate on physical channels is clearly not independent.
The easiest way to face this situation is to exploit for RSTAs a deterministic protocol overlay (e.g., HCCA or, possibly, solutions specifically engineered for seamless redundancy), which prevents them from transmitting on air at the same time.
This is not unfeasible, as these devices are usually part of the same control system.
Actual advantages of redundancy in presence of many nodes operating with random access is still an open issue, and is left as future work.
For such reason, the following analysis takes into account a unidirectional redundant link, denoted $\Link$, between two RSTAs (\emph{originating} RSTA and \emph{recipient} RSTA), and $\mathcal{C}_\Link = \left\{ A, B, ...\right\}$ is the set of the related physical channels.
To keep the testbed simple, only \emph{duplex} redundancy has been considered in the experiments, in which case $\Link$ is denoted $A\!+\!B$.
We assume that fragmentation, aggregation, and block acknowledgment mechanisms are not exploited, since they are not meant to meet the requirements of time-sensitive control applications.

\subsection{Seamless Redundancy Basics}
Although seamless redundancy also applies to unconfirmed transmissions, this paper only focuses on acknowledged exchanges.
Besides being practically mandatory to support reliable data transfers in Wi-Fi, the latter also enable effective optimizations when paired with seamless redundancy.
For our purposes, the term \emph{packet} denotes a unit of data to be exchanged over the network, as seen above the redundancy and data-link layers, and approximately corresponds to a MAC Service Data Unit (MSDU).
When a request is made in the originating RSTA to transmit a packet over a redundant link $\Link$, the related LRE issues concurrent requests on the adapters of its sub-STAs for sending identical \emph{copies} of the packet on every channel $C \in \mathcal{C}_\Link$.
Sub-STAs' operations are mostly disjoint \cite{TII_2017_PP_REDUNDANCY, 2017-TII-Guidelines}, and the transmission of each packet copy on the related adapter obeys MAC rules. 
This involves sending specific \emph{frames} on air, corresponding to MAC Protocol Data Units (MPDU), as well as dealing with carrier sensing, interframe spaces, and random backoff, issuing retransmissions upon missing acknowledgment, and so on.

Transmission of each packet copy on the related channel consists of one or more \emph{attempts}:
exactly one \emph{initial try} plus, upon unsuccessful delivery, a variable number of \emph{retries}.
To avoid starvation, a \emph{retry limit} is defined to bound the number of transmission attempts for any packet copy.
Each attempt is made up of a DATA frame (carrying user data), sent by the originator, immediately followed by an ACK frame (acknowledgment), returned by the unique recipient after a Short Interframe Space (SIFS).
Failed attempts are detected by the MAC in the originating sub-STA by setting a specific \emph{ACKtimeout} whenever a DATA frame is sent.
If the ACK frame is not received before the timeout expires, the attempt is considered as failed.

On the recipient RSTA, the LRE takes care of removing duplicates, reordering packets, and delivering them to the upper layers.
To do so, packet copies received on the physical channels of the redundant link are paired using specific sequence numbers added on the originating side.
For any packet, the DATA frame (either the initial try or a retry) that is correctly received first among all its copies is retained, while the following ones are discarded.

\subsection{Duplication Avoidance Mechanisms}
\label{sec:a}
In PoW, every packet is always sent on all channels.
Since channels can be assumed to be reasonably independent in a well-configured RBSS \cite{TII_2017_PP_REDUNDANCY}, overall spectrum consumption in PoW coincides with the sum of the traffics that would be induced if non-redundant, conventional \mbox{Wi-Fi} were used to send packets separately on each physical channel.
The main improvement \mbox{Wi-Red} \cite{2016-tii-WiRed} brings over PoW \cite{2012-WFCS-WoP1} are 
duplication avoidance mechanisms.
Basically, they are aimed at reducing the amount of frames sent on air.
This is a strict requirement when smooth coexistence has to be ensured with nearby networks, as the wireless spectrum is a scarce resource not to be wasted, especially in the $\unit[2.4]{GHz}$ band and, to a lesser (but increasing) extent, $\unit[5]{GHz}$ band.

Two kinds of DA mechanisms exist, namely \emph{reactive} and \emph{proactive}, both of which operate on the originating side of a redundant link.
The former simply prevents duplicate transmissions on air that are recognized to be useless, since an acknowledgment has already been received from the recipient, on any physical channel of the redundant link, for the information they convey.
Instead, the latter also includes heuristics aimed at complementing the reactive approach, so as to increase its effectiveness.

\section{Experimental Evaluation}
\label{sec:setup}
\begin{figure}[]
	\centering
	\includegraphics[width=1\columnwidth]{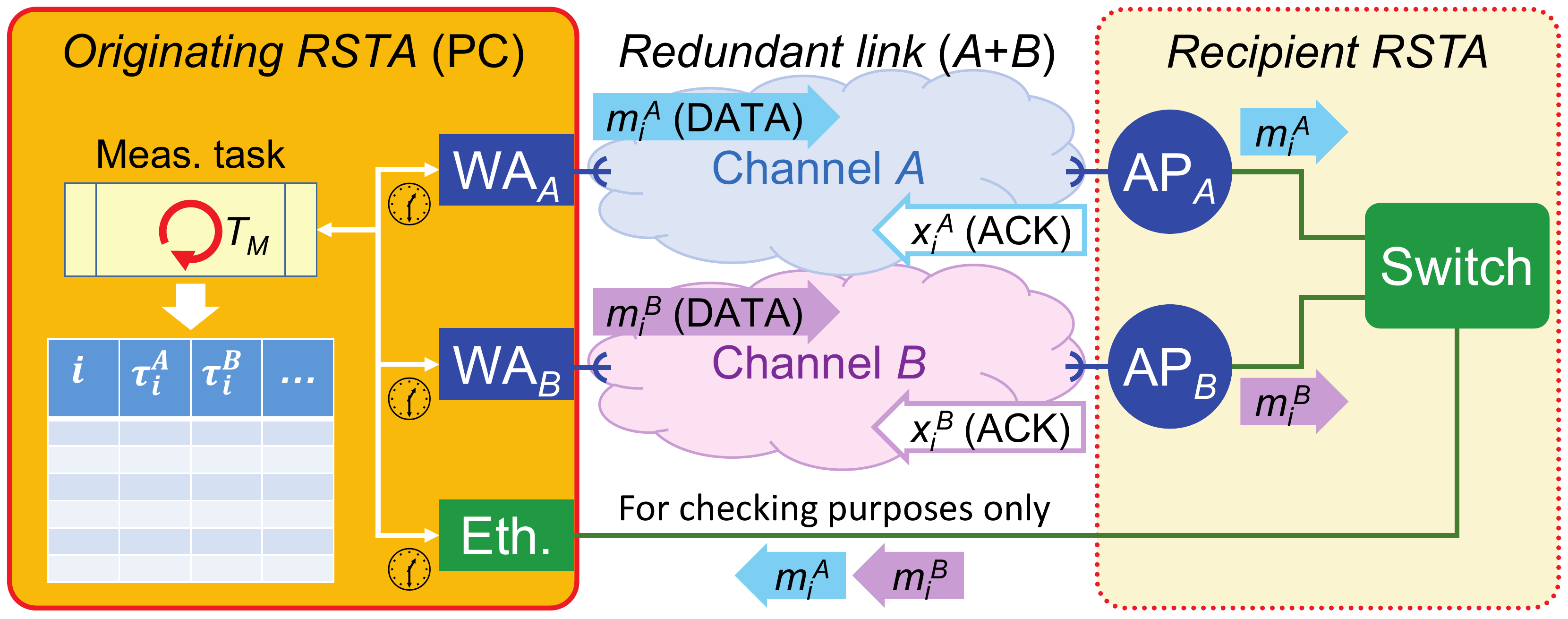}
	\caption{Experimental PoW testbed.}
	\label{fig:testbed}
\end{figure}

The experimental setup used for assessing Wi-Red performance resembles the one in \cite{2017-TII-Guidelines}.
However, the analysis carried out in this paper does not focus on communication quality, but rather on the ability of DA mechanisms to reduce spectrum consumption (which is the main performance metric here).
As shown in Fig.~\ref{fig:testbed}, the testbed is made up of a PC running the Linux OS (kernel v.~3.16.0) and provided with two identical Wi-Fi adapters (dual-band TP-Link TL-WDN4800), which emulates the behavior of the originating RSTA.
Two identical APs (NETGEAR WAC120), set to operate on completely disjoint frequencies to avoid interference (channel $1$ in the $\unit[2.4]{GHz}$ band and $44$ in the $\unit[5]{GHz}$ band), partially mimic the behavior of the recipient RSTA.
Each Wi-Fi adapter on the PC is associated to a distinct AP, which, on correct  reception of a DATA frame, returns an ACK frame to the related adapter and relays the packet to the PC Ethernet port through a Gigabit switch.

On the PC, a \emph{measurement task} produces a cyclic pattern of packets.
Generation period $T_M$ is set to $\unit[100]{ms}$. 
To obtain reliable statistics, each run lasted $24$ hours and included $N=864000$ packets, which are sent on the redundant link according to PoW rules (each packet is concurrently fed to both \mbox{Wi-Fi} adapters).
Timestamps are taken on every packet transmission and reception. 
Since they refer to the same time base, i.e., the CPU \emph{time stamp counter} (TSC) register of the PC, transmission latencies can be computed.
Following the approach in \cite{TII16-EDF}, receive times were not taken on packet arrival to the Ethernet port.
Instead, we modified Wi-Fi adapters' drivers so as to acquire a precise timestamp in kernel space upon ACK frame reception or \emph{ACKtimeout} expiry.
Doing so improves both accuracy and precision of measured transmission latencies.
In fact, although APs were, on average, quite fast in forwarding packets, they also introduced non-negligible jitters and delays.
This is no surprise, since they were not conceived for time-sensitive applications.
Results were checked against timestamps and losses obtained on the Ethernet port.

\subsection{Modeling Experimental Results}
\begin{table}
  \caption{Symbols and quantities used to characterize experiments.}
  \label{tab:sym}
  \centering
  \includegraphics[width=1\columnwidth]{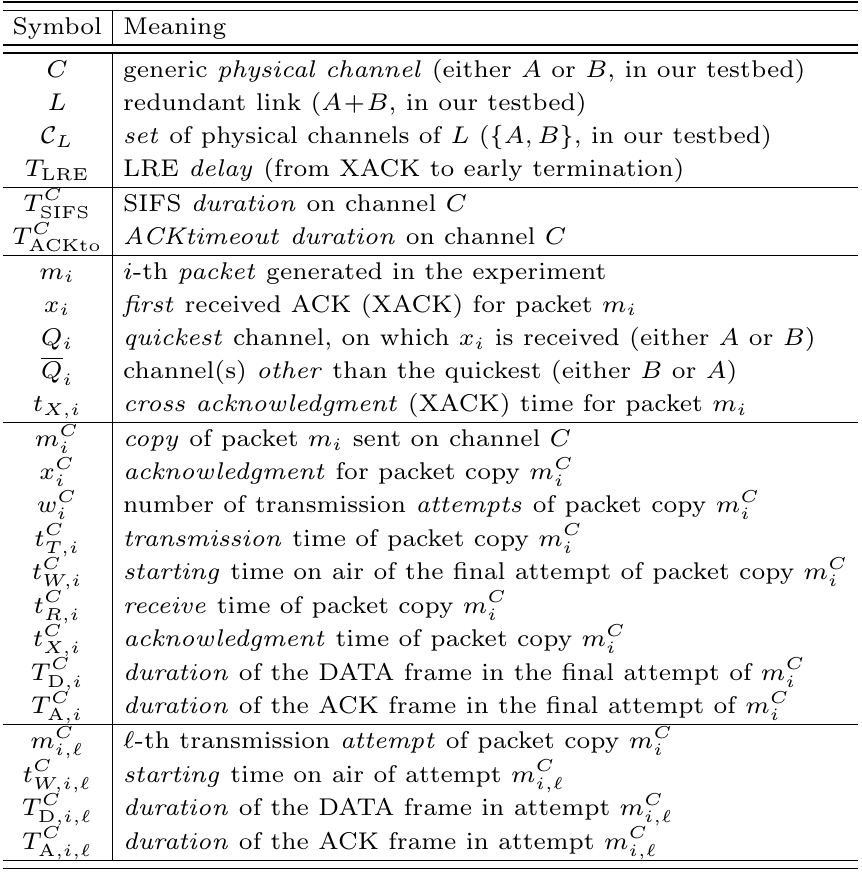}
\end{table}

Each experimental run in the testbed is modeled as a packet sequence $\mathcal{M}=\left(m_1, ...,m_N\right)$, where $m_i$ is the $i$-th packet generated by the measurement task.
For every packet, a transmission request is separately issued on all adapters.
The copy of $m_i$ sent on channel $\C \in \mathcal{C}_\Link$ (either $A$ or $B$, in our testbed) is denoted $m_i^C$, and its transmission outcome is described as a pair $\left\langle l^\C_i, d^\C_i \right\rangle$.
Boolean value $l^\C_i$ is the \emph{loss indication}, either $\operatorname{true}$ if $m_i$ went lost on $\C$ or $\operatorname{false}$ when it was successfully delivered.
In the following, Booleans are treated as integers, i.e., $\operatorname{true}\equiv 1$ and $\operatorname{false}\equiv 0$.
The non-negative value $d^\C_i$ is the \emph{transmission latency} of $m_i$ on $\C$, measured from the time $t^\C_{T,i}$ when the transmission request is issued on the related adapter to the time $t^\C_{R,i}$ when the packet is correctly received by the recipient, that is, $d^\C_i \triangleq t^\C_{R,i} - t^\C_{T,i}$.
Latency is only defined for packets for which $l^\C_i=0$, since on the contrary $t^\C_{R,i}$ does not exist. 

Behavior of redundant link $\Link$ can be analyzed starting from its physical channels.
Let $t_{T,i}^\Link$ be the time when packet $m_i$ is generated.
With the exception of proactive mechanisms \cite{2016-tii-WiRed}, described in Section~\ref{sec:PDA}, we can assume that, with very good approximation, transmission times for all the copies of any given packet coincide, that is $t_{T,i}^{C} \approx t_{T,i}^\Link,\forall C \in \mathcal{C}_\Link$.
Because of delays introduced at the MAC layer by the shared medium (due to, e.g., the channel sensed busy or retransmissions), the same does not necessarily hold for receive times $t_{R,i}^C$.

Links exploiting seamless redundancy satisfy two basic properties:
1) a packet $m_i$ is lost on $\Link$ (i.e., it never arrives to the recipient, not even after all allowed retries have been carried out by all adapters) only if its copies are definitely lost on every channel $C \in \mathcal{C}_\Link$;
2) the receive time of any packet $m_i$ correctly delivered on $\Link$ corresponds to the minimum between the receive times of its copies on all physical channels, i.e., $t_{R,i}^\Link=\operatorname{min}_{C \in \mathcal{C}_\Link, \l_i^C=0} \left( t_{R,i}^{C} \right)$.
This means that communication quality on a redundant link is never worse than on any of its physical channels.

\subsection{Measurement Technique}
From a practical viewpoint, two timestamps were acquired for every packet $m_i$ on each physical channel $C$: $t^\C_{T,i}$ is taken immediately before the transmission request is invoked, while $t^\C_{X,i}$ is taken in the device driver when transmission ends and passed to the measurement task through a character device.
In case of successful delivery, $t^\C_{X,i}$ is used to infer $t^\C_{R,i}$ and hence $d_i^C$.
Characterization of the redundant link $\Link$ is carried out a posteriori by applying PRP rules, that is 
$l^\Link_i = \prod_{C \in \mathcal{C}_\Link} l^C_i$ and, when $l^\Link_i=0$, $d^\Link_i \triangleq t^\Link_{R,i} - t^\Link_{T,i} \approx \operatorname{min}_{C \in \mathcal{C}_\Link, \l_i^C=0} \left( d^\C_{i} \right)$.

According to IEEE 802.11, retransmission of confirmed frames is automatically performed by Wi-Fi adapters upon missing acknowledgment.
Transmission attempts for the same packet on any channel $C \in \mathcal{C}_\Link$ are carried out consecutively, spaced by random backoff periods.
As soon as an ACK frame, denoted $x^\C_i$, is correctly received on $\C$ for packet copy $m^C_i$, its transmission on that channel ends.
Let $m^\C_{i,\ell}$ denote the $\ell$-th transmission attempt of packet $m_i$ on $C$, where $\ell=1$ indicates the initial try while $2 \leq \ell \leq w_i^C$ are retries ($w_i^C$ is the overall number of attempts performed for $m_i^C$).
From the originator viewpoint, attempts $m^\C_{i,\ell}$ with $\ell < w_i^C$ are certainly unsuccessful.
However, due to the retry limit, not necessarily the final attempt ($\ell = w_i^C$) succeeds.
Attempts performed on any physical channel $C$ in a run on the PoW testbed can be modeled as a sequence
\begin{equation}
	\label{eq:seqm}
	\left( 
		\underbrace{m^C_{1,1}, ..., m^C_{1,w_1^C}}_{\text{packet } m_1}, ... , 
		\underbrace{
			\overbrace{m^C_{i,1}}^{\begin{subarray}{c}\text{initial} \\ \text{try}\end{subarray}},
			\overbrace{m^C_{i,2}}^{\begin{subarray}{c}\text{$1$-st} \\ \text{retry}\end{subarray}}, ..., 
			\overbrace{m^C_{i,\ell}}^{\begin{subarray}{c}\text{$(\ell$-1)-th} \\ \text{retry}\end{subarray}}, ..., 
			\overbrace{m^C_{i,w_i^C}}^{\begin{subarray}{c}\text{final} \\ \text{attempt}\end{subarray}}, 
		}_{\text{packet } m_i}
		... \right)
\end{equation}
where each sub-sequence refers to a specific packet and includes all the related attempts.

The \emph{starting time} on air of $m^\C_{i,\ell}$, denoted $t^\C_{W,i,\ell}$, corresponds to the actual beginning of the $\ell$-th transmission attempt of $m_i$ on $C$.
Since carrier sensing is performed by Wi-Fi adapters independently, starting times of the initial attempts for packet $m_i$ on the different channels not necessarily coincide.
Even more so this holds for retries.
Attempts in \eqref{eq:seqm} are globally ordered according to their starting time on air, i.e.,
$t^\C_{W,i,\ell} < t^\C_{W,i,{\ell+1}}, 1 \leq i \leq N, 1 \leq \ell < w_i^C$ and $t^C_{W,i,w_i^C} < t^C_{W,i+1,1}, 1 \leq i < N$.
The overall \emph{duration} of attempt $m^\C_{i,\ell}$ includes three contributions, $T^{\C}_{\mathrm{D},i,\ell}$, $T^C_\mathrm{SIFS}$, and $T^C_{\mathrm{A},i,\ell}$, which correspond to the DATA frame, SIFS, and ACK frame, respectively.
Rate adaptation techniques in the originator directly affect $T^C_{\mathrm{D},i,\ell}$ and, indirectly, $T^C_{\mathrm{A},i,\ell}$ (by influencing the bit rate used by the recipient to reply).
Instead, $T^C_\mathrm{SIFS}$ is fixed and only depends on the specific Wi-Fi amendment, i.e., physical layer (PHY), selected for channel $C$.

\subsection{Assumptions on Channel Errors}
Both DATA and ACK frames may be corrupted during transmission on air.
If an error affects the ACK frame, the MAC on the recipient notifies the correct packet reception, but additional retries are attempted by the originator, until it either receives an ACK or exceeds the retry limit.
Typically, this causes the packet acknowledgment time on the originator to be delayed to one of the following retries.
If none of these succeed, the packet is considered lost by the originator, hence making its delivery state inconsistent with the recipient.
This issue is usually dealt with by recovery mechanisms in the upper protocol layers.

In order to simplify the following analysis, we will assume that errors always affect a transmission attempt as a whole, and not the ACK frame alone.
In the real world, this assumption is typically verified because:
1) ACK frames are very short and rely on robust modulation and coding schemes, and
2) under the reasonable hypothesis that the BSS does not suffer from the hidden node problem, they cannot incur in collisions not involving also the related DATA frame.
This means that, when a DATA frame is correctly transferred, very likely the related ACK frame also does.
We checked this assumption a posteriori on experimental samples, by comparing the set of packets arrived on the Ethernet port against those for which an ACK was received, and found that it actually holds in almost all the cases.
However, it is worth pointing out that the correct behavior of both seamless redundancy and DA mechanisms does not depend in any way on this assumption, which only leads to slightly pessimistic results for communication quality.

Under the above hypothesis, the events raised upon acknowledgment (on the originator) and reception (on the recipient) of packet $m_i$ on channel $\C$ refer to the same (successful) transmission attempt, which corresponds to $m^\C_{i,w_i^C}$.

\subsection{Limitations of Commercial Equipment}
Real Wi-Fi adapters, including those used in our testbed, typically raise interrupts only at the end of packet transmission and reception.
Thus, it is not possible to obtain timestamps $t^\C_{W,i,\ell}$ for inner attempts, but only for the final ones, in which case either an ACK frame has been correctly received (\emph{success}) or the \emph{ACKtimeout} has expired and the retry limit has been exceeded (\emph{failure}).
These conditions conclude the transmission process for packet copy $m_i^C$.
To keep notation simple, the starting time on air of the final attempt on $C$ is denoted $t^\C_{W,i}$, while the end of transmission coincides with $t^{\C}_{X,i}$.
The durations of the related DATA and ACK frames (indicated as $T^\C_{\mathrm{D},i}$ and $T^\C_{\mathrm{A},i}$, respectively) can be obtained by inspecting the driver.

In case of success ($l_i^C=0$) the starting time on air of the final transmission attempt of $m_i^C$ can be calculated as
\begin{equation}
	\label{eq:Wir}
	t^\C_{W,i} = t^\C_{X,i} - \left( T^\C_{\mathrm{D},i} + T^C_\mathrm{SIFS} + T^\C_{\mathrm{A},i} \right)
\end{equation}
and a reliable estimate for the receive time can be obtained as
\begin{equation}
	\label{eq:tR}
	t^C_{R,i} = t^C_{X,i} - \left( T^C_\mathrm{SIFS}+T^C_{\mathrm{A},i} \right),
\end{equation}
while in case of failure ($l_i^C=1$)
\begin{equation}
	\label{eq:Wirfail}
	t^{\C}_{W,i} = t^{\C}_{X,i} - \left( T^{\C}_{\mathrm{D},i}+T^{\C}_\mathrm{ACKto} \right) 
\end{equation}
where $T^{\C}_\mathrm{ACKto}$ is the \emph{ACKtimeout} duration on $\C$.

For the purpose of the following analysis, transmission of $m_i$ on physical channel $C$ of the testbed is completely described by a tuple $\tau^\C_i \triangleq \left\langle l^\C_i, t^\C_{T,i}, t^\C_{X,i}, w_i^{\C}, T^{\C}_{\mathrm{D},i}, T^{\C}_{\mathrm{A},i} \right\rangle.$
Overall, every experimental run is modeled as 
$\mathcal{T}=\left( \left\langle \tau^{A}_i, \tau^{B}_i \right\rangle \right)_{i=1...N}$.

\section{Reactive Duplicate Avoidance}
\label{sec:RDA}
Reactive Duplication Avoidance (RDA) is the basis on which all DA mechanisms rely.
Basically, the LRE of the originating RSTA exploits information provided at runtime by the MAC of its sub-STAs to prevent useless transmission attempts on air.
In particular, as soon as an ACK frame is received for a certain packet on any sub-STA, a \emph{cross ACK} event (XACK) is generated in the LRE, which forces the transmissions of the copies of that packet on all the other sub-STAs to be canceled.
Let $T_\mathrm{LRE}$ be the time taken by the LRE to force \emph{early termination}.
If RDA is implemented in the adapters' hardware/firmware, $T_\mathrm{LRE}$ can be as low as a few $\unit[]{\mu s}$, whereas when it is performed in software (e.g., in the drivers) $T_\mathrm{LRE}$ is expected to be in the order of several tens to hundreds $\unit[]{\mu s}$ and is noticeably less deterministic \cite{2017-WFCS-SDMAC}.
For packets whose transmission has failed on every sub-STA (and hence, on the redundant link), no XACK is generated, and hence no optimization is possible by RDA.

\begin{figure}[]
	\centering
	\includegraphics[width=1\columnwidth]{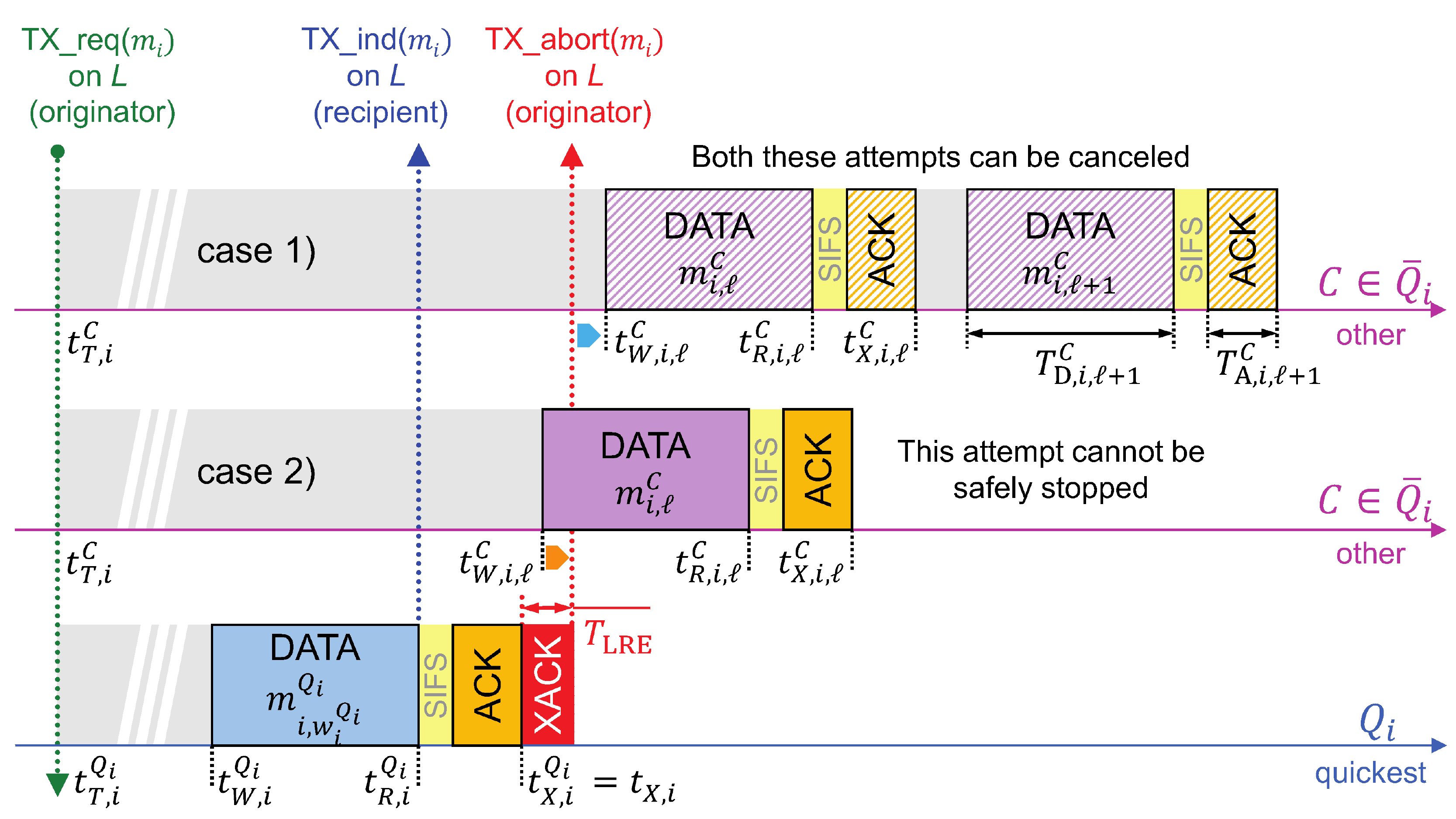}
	\caption{Early termination on XACK events carried out by RDA.}
	\label{fig:RDA}
\end{figure}

Concerning a packet $m_i$ successfully delivered on the redundant link $\Link$ (that is, for which an ACK frame is correctly received on at least one of its physical channels), let $\Qi$ be the \emph{quickest} channel, i.e., the one on which a valid ACK frame is received first (see the diagram at the bottom of Fig.~\ref{fig:RDA}).
In formulas, $\Qi = \operatorname{arg \, min}_{C \in \mathcal{C}_\Link, l_i^C=0} \left( t_{X,i}^C \right)$.
That ACK, as well as the XACK it triggers, are denoted $x_i$ and occur at time $t_{X,i} = t_{X,i}^{\Qi}$.
Channels other than $\Qi$ are collectively denoted $\Qni$, in formulas, $\Qni = \mathcal{C}_\Link \backslash \left\{ \Qi \right\}$.
For duplex redundancy (as in our testbed) one such channel exists.
With a slight abuse of notation, it can be directly identified as $\Qni$.

As shown in the topmost diagram of Fig.~\ref{fig:RDA} (case~1), according to RDA rules the LRE can prevent the $\ell$-th transmission attempt of $m_i$ on $C \in \Qni$ from starting provided that XACK processing has completed before $t^{C}_{W,i,\ell}$.
On the contrary (case~2), terminating the attempt midway has to be avoided, since this forces an error condition in all the receiving STAs (similarly to when a frame gets corrupted on air) and worsens coexistence with nearby Wi-Fi devices.
Moreover, since frame duration is set in advance in the frame preamble (encoded as data rate and length) and the channel is reserved up to the end of the ACK frame by the Network Allocation Vector (NAV), doing so would bring little benefits.
In formulas, attempt $m_{i,\ell}^{\C}$ (and the following) can be prevented on any channel $\C \in \Qni$ when
\begin{equation}
	\label{eq:term}
	t_{X,i} + T_\mathrm{LRE} < t^{\C}_{W,i,\ell} .
\end{equation}

\subsection{Performance Comparison between RDA and PoW}

\subsubsection{Communication Quality}
When the RBSS is correctly dimensioned, so that generated packets seldom suffer from appreciable queuing delays in local buffers, RDA offers the same performance as PoW.
In fact, RDA acts \emph{after} ACK reception, i.e., after the related packet has been delivered to the recipient.
This means that, in absence of queuing, RDA operations can not affect in any way neither latencies $d_i^{\Link}$ nor packet losses $l_i^{\Link}$ of PoW.
This is mostly the case of our experiments, as packets produced by the measurement task were intentionally spaced wide enough.
When traffic increases and the RBSS operates in saturated conditions, RDA sensibly improves communication quality over PoW \cite{2016-tii-WiRed}, because of its ability to lower network load, which reduces collisions and, hence, packet losses and mean latencies.
Latencies also shrink because transmission queues in the originating RSTA are drained more quickly.
Definitely, RDA adoption can never worse communication quality.

\subsubsection{Spectrum Consumption}
Unless interfering traffic and disturbance in the RBSS are very low (in which case redundancy is useless), RDA typically offers tangible advantages with respect to PoW, as our experiments indicate.
The ability to reduce wasted bandwidth improves further when network traffic increases and packets are buffered, since in this case they can also be removed from transmission queues.
Actual improvements in real scenarios heavily depend on the ability of the LRE to quickly terminate ongoing transmissions upon ACK reception.
As will be shown, delays introduced by, e.g., software LRE implementations, impair RDA performance. 
In order to reliably assess spectrum consumption on a real RDA testbed, modifications to the firmware of Wi-Fi adapters are likely required, which is unfeasible with most commercial devices (including ours).
For this reason, we evaluated bandwidth saving in RDA using a different approach.

\subsection{Virtual RDA Analysis on PoW Logs}
PoW behavior closely resembles \emph{basic} Wi-Red, i.e., seamless redundancy with no DA mechanisms.
Once started, transmission of each packet copy is carried out independently on the related channel as per IEEE 802.11 MAC rules.
Interestingly, experimental data acquired in a run from the PoW testbed can be used to analyze RDA performance as well.
In particular, the amount of transmission attempts that can be saved by early termination in the very same operating conditions (including environmental ones, which are almost impossible for us to replicate) can be inferred.
This provides a reliable estimate of the improvements RDA achieves over PoW in terms of spectrum consumption, without the need to actually implement RDA.

\begin{figure}[]
	\centering
	\includegraphics[width=1\columnwidth]{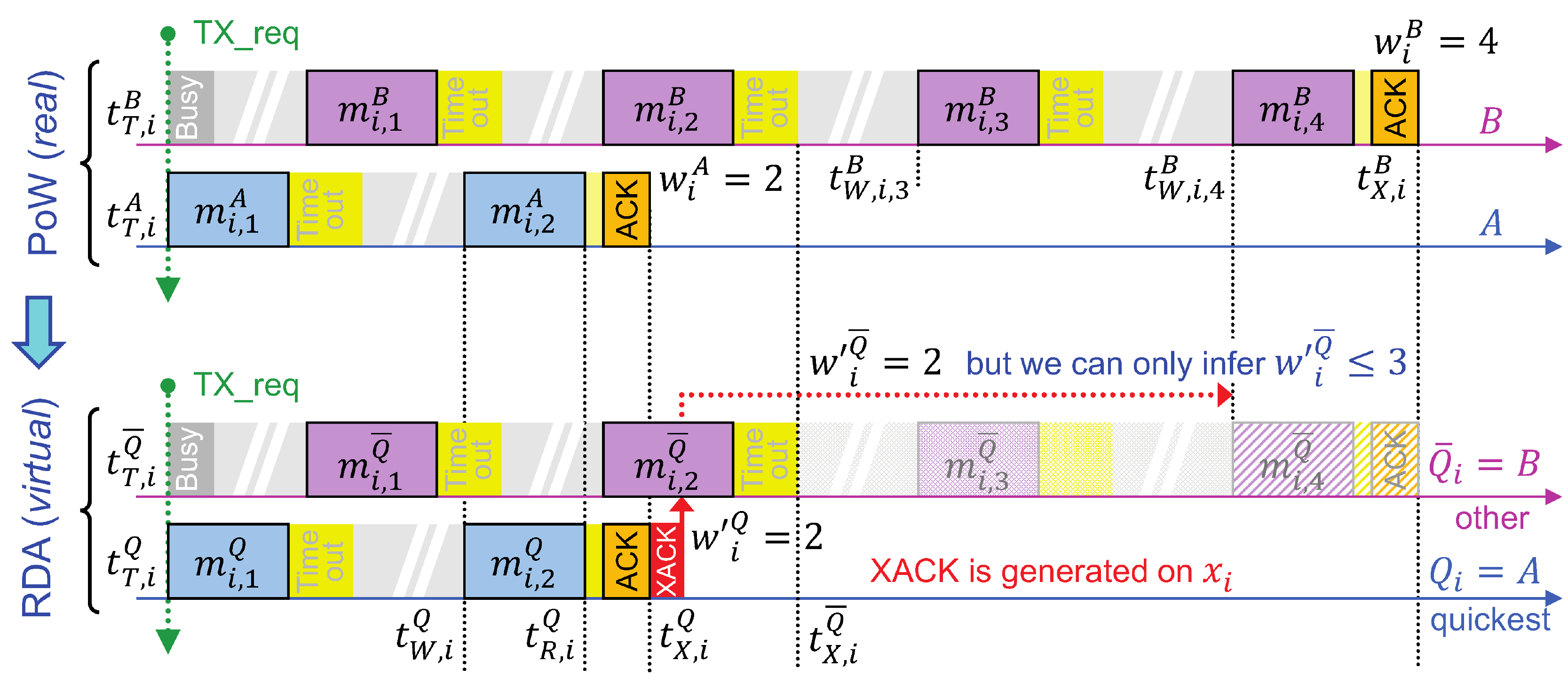}
	\caption{Virtual RDA experiment vs. real PoW experiment.} 
	\label{fig:virtual}
\end{figure}

A prerequisite for the following analysis is that, terminating a packet transmission in advance does not affect the following ones.
This is mostly true if, on every physical channel $C$, adjacent packet transmissions do not overlap, i.e., if transmission latency $d_i^\C$ never exceeds the generation period $T_M$.
Under such hypothesis, if a given PoW experiment was repeated, in \emph{exactly} the same operating and environmental conditions, using a real RDA testbed, the sequence of transmission attempts on air would simply consist of a subset of the PoW run.
In fact, some PoW attempts are possibly saved because of early termination.
In other words, for any attempt $m_{i,\ell}^C$ carried out when RDA is enabled there would be an equivalent attempt (same starting time on air and duration) carried out by PoW, but not vice-versa, because all attempts for which \eqref{eq:term} holds are prevented by RDA.
For instance, in the example shown in Fig.~\ref{fig:virtual}, RDA cancels the last two attempts carried out by PoW on $B \in \Qni$, that is, $m^{B}_{i,3}$ and $m^{B}_{i,4}$.

In the above analysis there is no need to define new quantities that explicitly refer to RDA, as they coincide with PoW.
The only difference lies in the number of transmission attempts, which for RDA is possibly smaller than PoW.
The number of attempts for packet $m_i$ on channel $\C$ in the RDA case is denoted $ {w'}_i^{C}$.
For the quickest channel ${w'}_i^{\Qi} = w_i^{\Qi}$ while, for any other channel $C \in \Qni$, $0 \leq {w'}_i^{\C} \leq w_i^{\C}$.
In case of early termination on $C \in \Qni$, the lowest value of $\ell$ for which \eqref{eq:term} is true  coincides with ${w'}_i^{\C}+1$.
In particular, ${w'}_i^{\C}=0$ refers to the case where also the initial attempt ($\ell = 1$) was prevented on $\C$ (which implies that $m_i$ was not sent there).
If \eqref{eq:term} is not verified for any $\ell \leq w_i^{\C}$, early termination is not possible for $m_i^{\C}$ and ${w'}_i^{\C} = w_i^{\C}$.

In our experiments on the PoW testbed, less than $0.1\%$ of the packets experienced queuing delays (see column $\Upsilon_{d>100\mathrm{ms}}^C$ in Table~\ref{tab:latency1}), i.e., more than $99.9\%$ of the packet transmissions did not overlap.
Therefore, above \emph{virtual} RDA analysis can be reliably applied to the related \emph{real} PoW logs.

\subsection{Experimental Evaluation of RDA}
Let $\Link$ be a redundant link like the one implemented by our PoW testbed, for which the hypotheses that permit to apply virtual RDA analysis hold.
In theory, early termination could be completely analyzed from the related experimental logs.
However, since starting times $t_{W,i,\ell}^C$ are only available for final attempts, exact ${w'}_i^{\C}$ values can not be determined.
Nevertheless, an indication can be still obtained on whether, for any given packet, some attempts can be saved.
This provides an upper bound on RDA spectrum consumption.
As shown in Fig.~\ref{fig:virtual}, if \eqref{eq:term} holds for the final attempt $m^{\C}_{i, \ell}$ on channel $C \in \Qni$, then \emph{at least one} attempt is prevented for sure on $C$.
In the sample packet transmission depicted in the figure, where ${w}_i^{B}=4$, this approach permits to infer that ${w'}_i^{B}\leq 3$ (although, actually, ${w'}_i^{B}=2$).

Let $e_i^C$ (\emph{early termination} condition) be a Boolean value that is true if and only if, owing to XACK $x_i$, RDA is able to terminate the transmission of packet $m_i$ on channel $\C \in \Qni$ in advance with respect to what is foreseen by conventional Wi-Fi.
As said before, this is only possible  for packets correctly delivered (and acknowledged) on $\Link$.
From \eqref{eq:term},
\begin{equation}
	\label{eq:EC1}
	e^{\C}_i \triangleq \left[ l_i^\Link = 0 \wedge t_{X,i}+T_\mathrm{LRE} < t^{\C}_{W,i} \right]
\end{equation}
where $t^{\C}_{W,i}$ is given by \eqref{eq:Wir} and \eqref{eq:Wirfail}, for successful and failed packet copy transmissions, respectively.
For the quickest channel, \eqref{eq:EC1} correctly provides $e_i^{\Qi} = 0$, so it holds for every $C \in \mathcal{C}_\Link$.
Above considerations identically apply when RDA is coupled with some heuristics (PDA).

\subsection{Metrics about Spectrum Consumption}
Reliability through time and frequency diversity comes to the price of a higher spectrum consumption.
The holistic approach provided by DA mechanisms offers non-negligible advantages in this respect.
The related benefits in terms of bandwidth  saving can be evaluated by analyzing the number of transmission attempts actually performed on air during any given experiment.

A number of simple metrics have been introduced to this extent.
In the following, when referring to redundant links, the kind of DA mechanism will be specified, if relevant, as superscript, in the place of the link.
Term DA generically refers to both RDA and PDA, while PoW denotes the lack of any DA.
In these cases, the link can be inferred from the context (either $\Link$, in general reasoning, or $A\!+\!B$, when referring to our experimental testbed).

\subsubsection{Saved Bandwidth}
The fraction of packets in an experimental run whose transmission is terminated early on physical channel $C \in \mathcal{C}_\Link$ is
\begin{equation}
	\label{eq:EC2}
	\overline{e}_{\operatorname{}}^{\C} \triangleq \frac{1}{N} \sum_{i=1...N} e_i^C .
\end{equation}

For any packet sent on the redundant link $\Link$, the average number of copies that experience early termination, all channels considered, is
\begin{equation}
	\label{eq:EC}
	\overline{e}_{\mathrm{}}^{\Link} \triangleq \frac{1}{N} \sum_{i=1...N} \sum_{C \in \mathcal{C}_\Link}  e_i^C = \sum_{C \in \mathcal{C}_\Link} \overline{e}^C.
\end{equation}
When $\Link ={A\!+\!B}$, as in our duplex testbed, there is only one channel other than the quickest, and $\overline{e}_{\mathrm{}}^{A\!+\!B}$ coincides with the fraction of packets terminated early on either $A$ or $B$.

Upon early termination of a packet copy, one or more transmission attempts are canceled for sure on the related channel.
This implies that $\overline{e}_{\mathrm{}}^{\Link}$ constitutes a \emph{lower bound} on the average number of useless attempts that, overall, can be prevented for each packet by RDA with respect to PoW, and provides an estimate of the bandwidth saving achieved over basic seamless redundancy.
Packets that are dropped on all physical channels but one are those for which the most savings are expected.
In fact, in the absence of early termination, all attempts up to the retry limit are (unsuccessfully) carried out on faulty channels.

Our testbed does not make $T^{\C}_{\mathrm{D},i}$ available when transmission on adapter $C$ fails, nor it can be easily inferred (because of rate adaptation mechanisms), and so $e_i^C$ can not be evaluated for the related packet copies.
This is not a severe drawback since, due to the effectiveness of MAC retransmissions, the fraction of packets that were dropped on either of the physical channels in our experiments was always very low (less than $0.0033\%$).
Hence, their contribution to $\overline{e}^\Link$ can be quietly neglected, by assuming for them $e_i^C=0$.
It is worth pointing out that the case when the loss rate is not negligible only leads to a pessimistic, smaller value for $\overline{e}^\Link$, which is acceptable as this quantity represents a lower bound.

\subsubsection{Network load}
The average number of transmission attempts per packet (including retries) carried out on any channel $\C$ during an experiment on the PoW testbed is
\begin{equation}
	\label{eq:WC}
	\overline{w}^{\C} \triangleq \frac{1}{N} \sum_{i=1...N} w_i^{\C} .
\end{equation}
As in Wi-Fi every packet is sent on air at least once, $\overline{w}^{\C} \geq 1$.

The value of $w_i^\C$ for successfully delivered packet copies can be obtained by inspecting the device driver of the related adapter,
whereas it is not available for failed copies.
In the latter case, using for $w_i^{\C}$ the default retry limit as per \mbox{Wi-Fi} specifications \cite{IEEE-STD_802-11_2016} is not the best choice, because of the rate adaptation mechanism (Minstrel) \cite{2013_ICC_Minstrel}, which changes transmission parameters dynamically according to some channel statistics.
Hence, for dropped packets (i.e., if $l_i^C = 1$) we assumed $w_i^{\C}$ to be equal to the maximum among all the values we measured in the experiments, that is $w_i^C = \operatorname{max}_{i'\in1...N\!,C'\in \mathcal{C}_\Link} \left( w_{i'}^{\C'} \right) = 21$.
Again, since the loss ratio in our experiments was quite low, this pessimistic approximation is perfectly acceptable.

Let $\overline{w}^{\Link}$ be the average number of transmission attempts on air per packet, considering all channels. 
It depends on the overhead caused by both retransmissions and seamless redundancy.
In PoW, every packet is always transmitted on all channels, so
\begin{align}
	\label{eq:WABPoW}
	\overline{w}^{\mathrm{PoW}} \triangleq \frac{1}{N} \sum_{i=1...N} \sum_{C \in \mathcal{C}_\Link}  w_i^C = \sum_{C \in \mathcal{C}_\Link} \overline{w}^C = \left| \mathcal{C}_\Link \right| \cdot \overline{w} ,
\end{align}
where $\overline{w} = \frac{1}{\left| \mathcal{C}_\Link \right|} \sum_{C \in \mathcal{C}_\Link} \overline{w}^C$ is the mean number of attempts per packet and per channel, and describes, on average, the behavior of a single conventional Wi-Fi channel.
Due to the different spectrum conditions of physical channels $C \in \mathcal{C}_\Link$, $w_i^C$ values usually do not coincide and the same holds for $\overline{w}^C$.
Since $\overline{w}^C \geq 1$, $\overline{w}^{\mathrm{PoW}} \geq \left| \mathcal{C}_\Link \right|$, where $\left| \mathcal{C}_\Link \right| = 2$ for duplex links.
Roughly speaking, spectrum consumption in our PoW testbed is, on average, twice as much as Wi-Fi.

When a DA mechanism is in effect, equation \eqref{eq:WABPoW} can be rewritten as
\begin{equation}
	\label{eq:WRDA}
	\overline{w}^\mathrm{DA} = \frac{1}{N} \sum_{i=1...N} \left( w_i^{\Qi} + \sum_{C \in \Qni} {w'}_i^{C} \right) .
\end{equation}

Unlike PoW, ${w'}_i^{\C}$ can be equal to $0$ for $C \in \Qni$, as the only constraint is that every packet has to be sent on air at least once on at least one channel.
Therefore, like plain Wi-Fi, $\overline{w}^{\mathrm{DA}}$ could be as low as $1$.
This happens when, for every packet $m_i$, ${w}_i^{\Qi}=1$ and ${w'}_i^{\Qni}=0$.
In theory, for some specific runs, spectrum consumption for $\Link$ may be even lower than any of its physical channels when used as simplex links.
In other words, condition $\overline{w}^{\mathrm{DA}} < \operatorname{min}_{C \in \mathcal{C}_\Link} \left( \overline{w}^{C} \right)$ might possibly hold.
For instance, when ${\Link}={A\!+\!B}$, consider the case $w_i^{\Qi} < w_i^{\Qni}$ and ${w'}_i^{\Qni}=0$ (i.e., the number of attempts that would be performed on $\Qni$ in the absence of DA is higher than on $\Qi$, but all of them are prevented by DA), where $\Qi=A$ if $i$ is even and $\Qi=B$ otherwise.

In general, $e_i^{\C} = 0 \rightarrow {w'}_i^{\C} = w_i^{\C}$, while $e_i^{\C} = 1 \rightarrow {w'}_i^{\C} < w_i^{\C}$.
This implies that ${w'}_i^{\C} \leq w_i^{\C} - e_i^{\C}$.
For a given DA mechanism, from \eqref{eq:EC}, \eqref{eq:WABPoW}, and \eqref{eq:WRDA}, we have
\begin{equation}
	\overline{w}^{\mathrm{DA}} \leq \overline{w}^{\mathrm{PoW}} - \overline{e}^{\mathrm{DA}}.
\end{equation}
Hence, DA mechanisms never worsens spectrum consumption of Wi-Red.

\subsubsection{Communication efficiency} 
from the originator viewpoint, it is defined as the reciprocal of the average number of transmission attempts per packet.
For physical channel $C$ it corresponds to $\eta^{\C} = 1/\overline{w}^{\C}$.
While being related to the \emph{goodput}, they do not coincide, as $\eta^{\C}$ does not take into account protocol overheads (frame preambles, MAC headers and trailers, acknowledgments, inter-frame spaces, random backoff, etc.), but only retransmissions.
Definition for redundant link $\Link$ is similar, $\eta^{\Link} = 1/\overline{w}^{\Link}$.
If DA is used, a lower bound can be found for $\eta^\mathrm{DA}$, defined as 
\begin{equation}
	\check{\eta}^\mathrm{DA} \triangleq \frac{1}{\overline{w}^\mathrm{PoW} - \overline{e}^\mathrm{DA}} \leq \eta^\mathrm{DA}.
\end{equation}

\subsubsection{Relative network load}
The load on the link for a given DA mechanism with respect to PoW is defined as
\begin{equation}
	\label{eq:vtheta}
	\vartheta^{\mathrm{DA}} \triangleq \frac{\overline{w}^{\mathrm{DA}}}{\overline{w}^{\mathrm{PoW}}} =	\frac{\overline{w}^{\mathrm{DA}}}{\left| \mathcal{C}_\Link \right| \cdot \overline{w}} 
\end{equation}
where $1/\overline{w}^{\mathrm{PoW}} \leq \vartheta^{\mathrm{DA}} \leq 1$ (the lower, the better).
Similarly, the relative load implied, on average, over plain, simplex Wi-Fi links, is expressed as $\Theta^{\mathrm{DA}} \triangleq {\overline{w}^{\mathrm{DA}}} / {\overline{w}} = \left| \mathcal{C}_\Link \right| \cdot \vartheta^{\mathrm{DA}}$.

Equation \eqref{eq:vtheta} can be rewritten as $\overline{w}^{\mathrm{DA}} = \vartheta^{\mathrm{DA}} \cdot \left| \mathcal{C}_\Link \right| \cdot \overline{w}$, which shows that the average number of transmission attempts on air per packet, including all channels, is given by the product of three factors:
the relative load of the specific DA technique over PoW, the number of channels, and the mean per-packet load in Wi-Fi due to retransmissions, obtained by averaging all channels of the redundant link.
Only the first factor depends on DA effectiveness (which is affected, e.g., by $T_\mathrm{LRE}$).
Thus, $\vartheta^{\mathrm{DA}}$ is a valuable metric to assess the ability of a certain DA mechanism to save bandwidth.

Due to the limitations of our testbed, what we can measure is an upper bound on $\vartheta^{\mathrm{DA}}$, i.e.,
\begin{equation}
	\hat{\vartheta}^{\mathrm{DA}} \triangleq \left( 1 - \frac{\overline{e}^{\mathrm{DA}}} {\overline{w}^{\mathrm{PoW}}} \right)	\geq \vartheta^{\mathrm{DA}}.
\end{equation}
Therefore, DA performance is never worse than what we found using our approximate analysis.

\subsubsection{Simplex transmissions on the redundant link}
The last quantity we consider is the fraction of packets which, as a consequence of early termination, are only sent on $\Qi$, that is, those for which ${w'}_i^{C}=0, \forall C \in \Qni$.
From the point of view of nearby wireless STAs, transmission of these packets appears to take place as on a conventional Wi-Fi (\emph{simplex}) link, which is not fixed but dynamically selected among channels $C \in \mathcal{C}_\Link$.

Counting these packets exactly is not possible in our testbed.
However, condition 
\begin{equation}
	\label{eq:ZiC}
	z_i^C \triangleq \left[ e_i^C \wedge \left( w_i^{C}=1 \right) \right]
\end{equation}
implies that no transmission attempts at all are performed for $m_i$ on $C$.
In fact, $w_i^{C}=1$ means that only the initial attempt is performed by PoW, but it is canceled by DA since $e_i^C=1$.
For the quickest channel \eqref{eq:ZiC} correctly yields $z_i^{\Qi}=0$.
Hence, $\overline{z}^{C} = \frac{1}{N} \sum_{i=1...N} z_i^{\C}$ is a lower bound on the fraction of packets whose transmission is fully prevented on $C \in \mathcal{C}_\Link$ by DA.

For the redundant link $\Link$, Boolean condition $z_i^{\Link} \triangleq \prod_{C \in \Qni} z_i^C$ implies that $m_i$ transmission is completely prevented on every channel other than the quickest.
We are interested in the metric
\begin{equation}
	\overline{z}^{\Link} = \frac{1}{N} \sum_{i=1...N} z_i^{\Link},
\end{equation}
which represents a lower bound on the fraction of packets that were sent on air on one channel only (the quickest), as on a simplex link. 
Since in our duplex testbed there is only one channel other than $\Qi$, $z_i^A$ and $z_i^B$ can not be both $\operatorname{true}$, which means that $z_i^{A\!+\!B} = z_i^A + z_i^B$.
In this specific case, $\overline{z}^{A\!+\!B}= \overline{z}^{A} + \overline{z}^{B}$.
Given that $z_i^{A\!+\!B} \rightarrow \left( e_i^{A\!+\!B}=1 \right)$, where $e_i^{A\!+\!B} = e_i^{A} + e_i^{B}$, it also follows that $\overline{z}^{A\!+\!B} \leq \overline{e}^{A\!+\!B}$.

\subsection{Experimental Results for RDA}
To assess the improvements RDA can bring in the real world, three experiments were run on the PoW testbed in different operating conditions.
Channel $A$, set in the $\unit[2.4]{GHz}$ band, was quite crammed, due to the presence of many active wireless devices in nearby premises, which were not under our control.
To try ensuring some repeatability, every experiment lasted for a whole day (24 hours).
Even so, results concerning channel $A$ in the different runs can not be reliably compared one another.
Channel $B$, set in the $\unit[5]{GHz}$ band, had very light load. 
Thus, random interfering traffic was purposely injected on that channel during experiments.
Interference consisted in the repeated generation of packet bursts.
Payload size of interfering packets was set to $\unit[1500]{bytes}$.
The number of packets in each burst was not fixed, but followed an exponential random distribution with mean value $300$ and truncated to $1500$.
Within any burst, transmission requests for packets were spaced by $\unit[400]{\mu s}$.
The duration of the gap between the end of a burst and the beginning of the following one was randomly selected, according to an exponential distribution with mean value $\unit[200]{ms}$ and truncated to $\unit[20]{s}$.

\begin{table*}[t]
  \caption{Experimental results for physical channels $A$ and $B$ 
	and the redundant link $A\!+\!B$ in RDA ($T_\mathrm{LRE}=\unit[0]{\mu s}$) .}
      \label{tab:latency1}
      \centering
      \includegraphics[width=2\columnwidth]{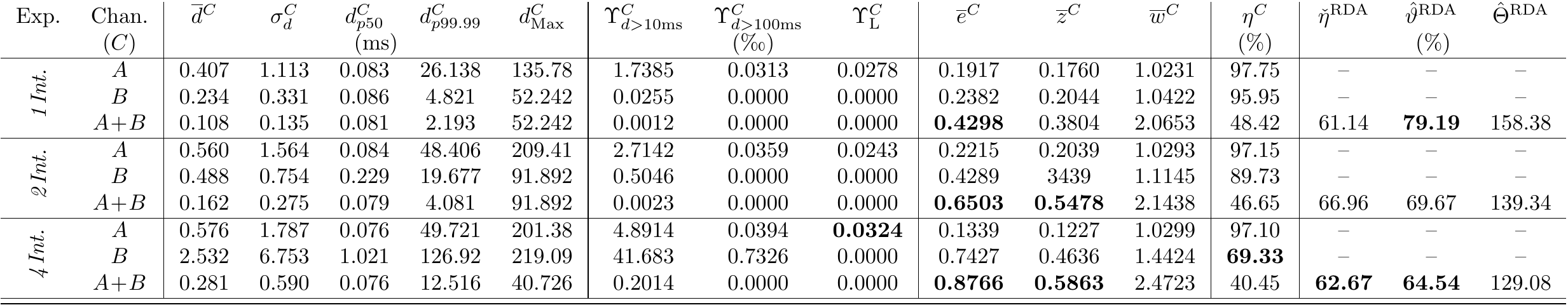}
\end{table*}

Runs were carried out with either $1$, $2$, or $4$ interfering STAs on $B$ (\emph{1\,Int.}, \emph{2\,Int.}, and \emph{4\,Int.}, respectively).
Experimental results, including bounds calculated using virtual RDA analysis, are reported in Table~\ref{tab:latency1}, under the assumption that $T_{\mathrm{LRE}}=0$ (i.e., when early termination is assumed to be implemented in hardware in the RSTA).
For each experiment, three rows are included, for physical channels ($A$, $B$) and the redundant link (\mbox{$A\!+\!B$}).
The first set of columns includes statistics about the transmission latency $d$: mean value, standard deviation, median, $99.99$ percentile, and maximum.
In the second set of columns, the \emph{deadline miss ratio} $\Upsilon_{d>H}^C$ (percentage of packets whose latency exceeded $H$, where $H$ is set to either $10$ or $\unit[100]{ms}$) and the \emph{packet loss ratio} $\Upsilon_{\mathrm{L}}^C$ are shown.
Although this paper does not focus on communication quality, these quantities provide a glimpse of the improvements seamless redundancy achieves.

The most interesting results, reported in the sets of columns on the right, concern RDA ability to prevent useless attempts.
Relevant measured quantities for physical channels $A$ and $B$ and the redundant link $A\!+\!B$ include the fractions of early terminated packets ($\overline{e}^{\C}$) and simplex packets ($\overline{z}^{\C}$), as well as the mean number of transmission attempts performed per packet ($\overline{w}^C$).
Communication efficiency when no DA is in use, both for physical channels ($\eta^{A}$ and $\eta^{B}$) and the redundant link ($\eta^{\operatorname{PoW}}$), is given in column $\eta^C$.
Finally, the rightmost columns report metrics about RDA computed using virtual analysis, i.e., the lower bound on communication efficiency ($\check{\eta}^{\operatorname{RDA}}$) and the upper bounds on the relative network load with respect to either PoW ($\hat{\vartheta}^{\mathrm{RDA}}$) or plain Wi-Fi ($\hat{\Theta}^{\mathrm{RDA}}$).

Experimental results are quite valuable and show that, in real operating conditions, RDA cuts redundancy overhead tangibly.
The lower bound $\check{\eta}^{\operatorname{RDA}}$ on communication efficiency (and hence, $\eta^{\operatorname{RDA}}$ as well) was always noticeably higher than $\eta^{\operatorname{PoW}}$, irrespective of the interfering traffic, and the upper bound $\hat{\vartheta}^{\mathrm{RDA}}$ on the network load relative to PoW ranged from $64.54\%$ to $79.19\%$.
This means that all benefits on communication quality brought by seamless redundancy are achieved at a fraction of the bandwidth.
Since spectrum conditions may vary suddenly and unexpectedly, fair comparison between redundant and simplex \mbox{Wi-Fi} links requires that the worst of the physical channels is taken into account.
As can be seen in the table, in the \emph{4\,Int.} case communication efficiency of RDA is marginally worse than channel $B$ alone ($62.67\%$ vs. $69.33\%$), but communication quality on $A\!+\!B$ is much better than both $A$ and $B$.

An additional benefit of RDA is that, the higher the overall network traffic, the better it works.
See, e.g., when interfering traffic on $B$ is varied.
The measured fraction $\overline{e}^{\mathrm{RDA}}$ of early terminated packet transmissions with $1$, $2$, and $4$ interferers was equal to $42.98\%$ (\emph{1\,Int.}), $65.03\%$ (\emph{2\,Int.}), and $87.66\%$ (\emph{4\,Int.}), respectively.
Concerning the fraction $\overline{z}^{\mathrm{RDA}}$ of simplex packet transmissions, when traffic is not negligible more than half of the packets ($54.78\%$ and $58.63\%$, for \emph{2\,Int.} and \emph{4\,Int.}, respectively) were actually sent only on one channel, hence causing no overhead on the other.
This is no surprise, as RDA effectiveness improves when variability of transmission latencies on replicated channels gets higher.
Such behavior is highly desirable since, when overall network traffic increases, the overhead due to redundancy consequently shrinks, hence lowering the impact of transmitting RSTAs on the network load.
This counteracts congestion phenomena affecting random access techniques, which worsen the quality of communication (see, e.g., the latency on channel $B$ when interfering traffic is increased).
As a consequence, network stability (overall traffic on air vs. total offered load) improves.

\begin{figure*}
  \centering
  \includegraphics[width=2\columnwidth]{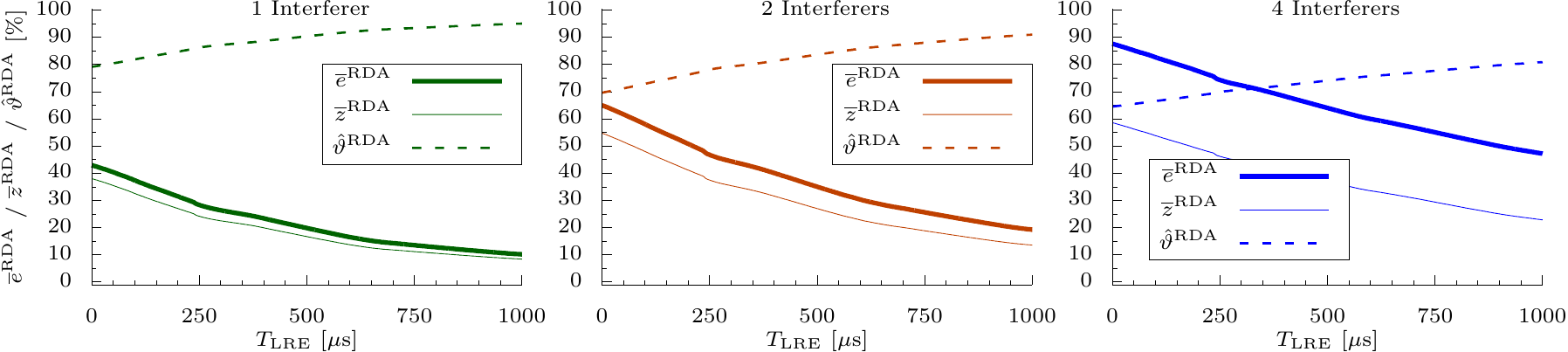}
  \caption{Fractions of early terminated ($\overline{e}^{\mathrm{RDA}}$) and simplex ($\overline{z}^{\mathrm{RDA}}$) transmissions, and relative load ($\hat\vartheta^{\mathrm{RDA}}$), vs. $T_\mathrm{LRE}$ (for RDA).}
  \label{fig:TLRE}
\end{figure*}

The effect of using sub-optimal (e.g., software) RDA solutions, modeled by considering non-negligible LRE latencies, is shown in Fig.~\ref{fig:TLRE}.
Three plots are reported, corresponding to the above experimental runs, where $T_\mathrm{LRE}$ in \eqref{eq:EC1} is virtually varied in the range $\left[0-1000\right]\unit[]{\mu s}$.
As can be seen, delays due to XACK processing always worsen spectrum consumption as, on average, less transmission attempts can be terminated in advance.
For example, in the \emph{1\,Int.} case, $\overline{e}^{\mathrm{RDA}}$ decreases by a factor $4$ (from $42.98\%$ to $10.18\%$) when $T_\mathrm{LRE}$ is increased to $\unit[1]{ms}$.
For this reason, implementation of RDA in the hardware (or firmware) of Wi-Fi adapters is highly suggested.
This also confirms the advantages brought by our approximate virtual approach to RDA evaluation (based on the analysis of PoW experimental logs) with respect to measurements carried out directly on a non-properly optimized software RDA implementation. 
Moreover, it is important to remark the ability of the virtual approach to infer results for different techniques and parameters starting from the same set of data, allowing for a fair comparison between them.

\section{Proactive Duplication Avoidance}
\label{sec:PDA}
Proactive Duplication Avoidance (PDA) is meant to improve RDA behavior by further reducing the number of useless duplicates sent on air.
Unlike RDA, which has no counter-indications, PDA may set some trade-off among performance indices (e.g., the approach described below trades responsiveness for communication efficiency).
Therefore, when employing these mechanisms, a compromise has to be necessarily found, typically by tuning suitable configuration parameters.

PDA often relies on heuristics, which could be arbitrarily complex and possibly intertwined with the IEEE 802.11 MAC.
The class of PDA mechanisms we considered in this paper only concerns duplex redundant links, and operates by intentionally displacing transmission requests on physical channels.
In this way, the likelihood that the LRE succeeds in canceling the transmission of the deferred copy may increase tangibly, hence reducing spectrum consumption.

\subsection{Timed Duplicate Deferral}
The PDA technique we analyze here is called Timed Duplicate Deferral (TDD).
A similar solution, not coupled with early termination, was proposed in \cite{2013_WCCIT_TDD}.
TDD is much simpler than the Dynamic Duplicate Deferral (DDD) technique presented in \cite{2016-tii-WiRed}, as its operation does not rely on any indication provided at runtime by the MAC layer.
Thus, it can be easily implemented using commercial Wi-Fi adapters.

TDD operates as follows: upon generation of packet $m_i$, the LRE immediately invokes its transmission on $A$ (referred to in the following as \emph{primary} channel).
At the same time, it starts a countdown timer initialized to the \emph{deferral time} $T_{D,i}$.
When $T_{D,i}$ expires, and provided that the XACK from $A$ has not been raised in the meanwhile, transmission is also invoked on $B$ (\emph{secondary} channel).
In formulas
\begin{equation}
	\label{eq:TT}
	t^B_{T,i} = t^A_{T,i} + T_{D,i} .
\end{equation} 
A distinct $T_{D,i}$ value can be chosen for each packet $m_i$ depending, e.g., on its payload size and the bit rate chosen at runtime by the rate adaptation mechanism.

\subsection{Experimental Evaluation of TDD}
TDD performance can be experimentally assessed in our testbed using two different approaches, which will be shown to provide very similar results.
In both cases, bandwidth saving is evaluated when varying the deferral time, by analyzing logs in the same way as for RDA.

\subsubsection{Real transmission deferral}
\label{sec:rtdd}
The most straightforward way to analyze TDD behavior is to rely on a slightly modified PoW testbed, where the transmission request issued by the measurement task on $B$ for each packet $m_i$ is actually delayed by $T_{D,i}$ with respect to $A$.
To make timing accuracy independent from the implementation of the measurement task and the underlying computing platform (including the operating system), the exact displacement between the copies of every packet $m_i$ (that in theory corresponds to $T_{D,i}$) was evaluated from measured timestamps as $t_{T,i}^B - t_{T,i}^A$.

\subsubsection{Virtual transmission deferral}
\label{sec:vtdd}
Approximate yet satisfactory results can be obtained from the logs acquired on the unmodified PoW testbed (i.e., without any displacement between $t_{T,i}^A$ and $t_{T,i}^B$).
Results in \cite{TII_2017_PP_REDUNDANCY} show that, in a well-designed redundant link, channel conditions can be considered statistically independent.
This means that deferring transmission requests on $B$ does not affect the behavior of $A$, and vice-versa.
Moreover, channel behavior in the short term is mostly stationary, hence, statistics on experimental results do not change upon displacement in time, as long as the offset is short enough (in the order of $\unit[1]{ms}$).

According to the above hypotheses, one can reasonably assume that the sequence of samples obtained by:
a) adding the deferral time to every timestamp ($t^B_{T,i}$ and $t^B_{X,i}$) in the tuples $\tau^{B}_i, {i=1...N}$ of a log acquired from channel $B$ in the unmodified PoW testbed, 
and b) leaving all the other quantities ($l^B_i$, $w_i^{B}$, $T^{B}_{\mathrm{D},i}$, and $T^{B}_{\mathrm{A},i}$) unmodified, could satisfactorily mimic the sequence of tuples acquired on $B$ in a real setup like the one described in Section~\ref{sec:rtdd}, where transmissions of copies are actually displaced.
This approach permits to create a bunch of \emph{virtual} TDD experiments from a single \emph{real} PoW experimental run, where $T_{D,i}$ values (provided that they are small, so as not to contradict stationary hypothesis) can be decided during post-analysis.

In the following, timestamps with the double prime symbol refer explicitly to virtual TDD experiments.
For example, the estimated ACK reception time for packet $m_i$ on the secondary channel $B$ in a virtual TDD experiment (characterized by the sequence of $T_{D,i}$ values) is
\begin{equation}
	\label{eq:TRB1}
	t''^B_{X,i} = t^B_{X,i} + T_{D,i}
\end{equation} 
where $t^B_{X,i}$ is the timestamp taken in the unmodified PoW testbed.
Timestamps $t''^B_{T,i}$ and $t''^B_{W,i}$ are defined similarly.
For the primary channel $A$, virtual and measured timestamps coincide.

In presence of TDD deferral, the transmission latency on the redundant link for packet $m_i$ is $d^{\mathrm{TDD}}_i = \operatorname{min}\left( d^A_{i}, d^B_{i} + T_{D,i}\right)$.
This implies that, unlike RDA, $d^{\mathrm{TDD}}_i\geq d^{\mathrm{PoW}}_i$.

The same approach used for RDA analysis can be applied to estimate the effects of early termination in TDD, by replacing real samples in the log with virtually deferred ones.
Since channel usage is no longer symmetric, \eqref{eq:EC1} splits into
\begin{align}
\label{eq:eABiprime}
	{e}_i^{B} \triangleq \left[ l_i^A = 0 \wedge {t''}^{A}_{X,i} + T_\mathrm{LRE} < {t''}^{B}_{W,i} \right] \\ \nonumber
	{e}_i^{A} \triangleq \left[ l_i^B = 0 \wedge {t''}^{B}_{X,i} + T_\mathrm{LRE} < {t''}^{A}_{W,i} \right]
\end{align}
and, from \eqref{eq:TRB1},
\begin{align}
\label{eq:eABi}
	{e}_i^{B} \triangleq \left[ l_i^A = 0 \wedge t^{A}_{X,i} + T_\mathrm{LRE} < t^{B}_{W,i}+ T_{D,i} \right] \\ \nonumber
	{e}_i^{A} \triangleq \left[ l_i^B = 0 \wedge t^{B}_{X,i} + T_{D,i} + T_\mathrm{LRE} < t^{A}_{W,i} \right].
\end{align}

\subsection{Validation of Virtual TDD Analysis}
\label{sec:validation}
A preliminary experiment was carried out to assess if results obtained through virtual deferral of samples performed offline are comparable to actually displacing transmission requests in the testbed at runtime.
Since the payload size of packets sent by the measurement task was fixed, all $T_{D,i}$ offsets in experimental runs have to be set to the same value $T_{D}$.
Jitter induced by the operating system on the actual displacement between the transmission times of the two copies of each packet in the modified PoW testbed described in Section~\ref{sec:rtdd} was found to be quite small. 
Therefore, we can practically assume that the deferral time in any run is fixed and coincides with the mean value of the measured offset, that is $T_{D} \approx \overline{T}_{D} = \frac{1}{N} \sum_{i=1...N}\left( t_{T,i}^B - t_{T,i}^A \right)$.

We performed several runs, where $T_{D}$ was varied in the range $\left[ -250 ... +250\right]\unit[]{\mu s}$.
The nominal displacement used in each run was selected in such a way that the related $\overline{T}_{D}$ approximately corresponds to one among $17$ predefined values.
In the case $T_{D}=0$ (which corresponds to pure RDA), transmissions were not displaced, as in the unmodified PoW testbed.
Negative $T_D$ values refer to experiments where $B$ was the primary channel, while $A$ was deferred by $\left|T_D\right|$.
These cases are explicitly included because behavior of channels is not symmetric.

\begin{figure}[t]
	\centering
	\includegraphics[width=1\columnwidth]{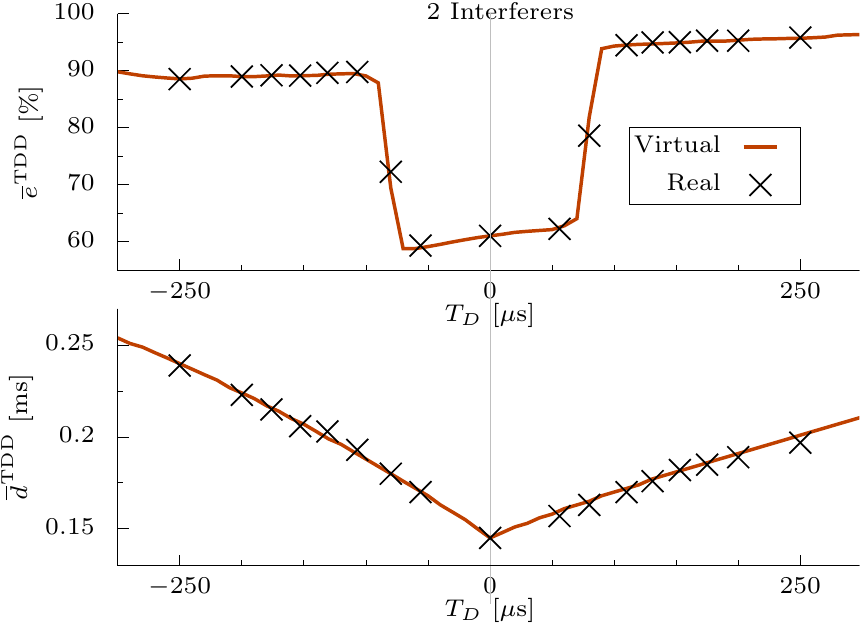}
        \caption{Comparison of $\overline{e}^{\mathrm{TDD}}$ and $\overline{d}^{\mathrm{TDD}}$ obtained with real and virtual transmission deferrals.}
      \label{fig:size}
\end{figure}

To mitigate discrepancies due to slow variations of the wireless spectrum conditions, runs for the different deferral times were interleaved in a single experiment, which lasted one day. 
Such a kind of assessment was lacking in \cite{2017-WFCS-Dupl}.
Two interferers were activated on $B$ during the experiment (as in the \emph{2\,Int.} case).
Results, concerning both the fraction of early terminated packets ($\overline{e}^{\operatorname{TDD}}$) and mean transmission latency ($\overline{d}^{\operatorname{TDD}}$), are reported in Fig.~\ref{fig:size} using ``$\times$'' symbols.

The very same logs, restricted to the subset of samples for which $T_D=0$ (i.e., when the testbed operated according to pure PoW, with no deferrals), were then used to generate virtual TDD runs according to Section~\ref{sec:vtdd}, where $T_D$ was varied with fine granularity in the range $\left[ -300 ... +300 \right] \mu s$.
Resulting plots are superposed to Fig.~\ref{fig:size} using solid lines.
As can be seen, there is a very good agreement between the results obtained with the two approaches.
This means that virtual deferral carried out during post-analysis provides a reliable approximation of real deferral performed by TDD.
For this reason, the relation between spectrum consumption, communication quality, and $T_D$, will be analyzed in the following by using the virtual approach.

\subsection{Experimental Results for TDD}
To provide coherent results, the very same experimental PoW logs used in Section~\ref{sec:RDA} to analyze RDA (\emph{1\,Int}., \emph{2\,Int.}, and \emph{4\,Int.}, depending on the amount of interference on $B$) were employed to carry out virtual TDD analysis.
First, they were processed to cope (virtually) with transmission deferral; then, the same analysis as for RDA was performed to estimate the effects of early terminations.
In particular, we analyzed the tradeoff between spectrum consumption and communication quality when $T_D$ is varied between $0$ (that corresponds to pure RDA) and $\unit[500]{\mu s}$.

\begin{figure*}
  \centering
  \includegraphics[width=2\columnwidth]{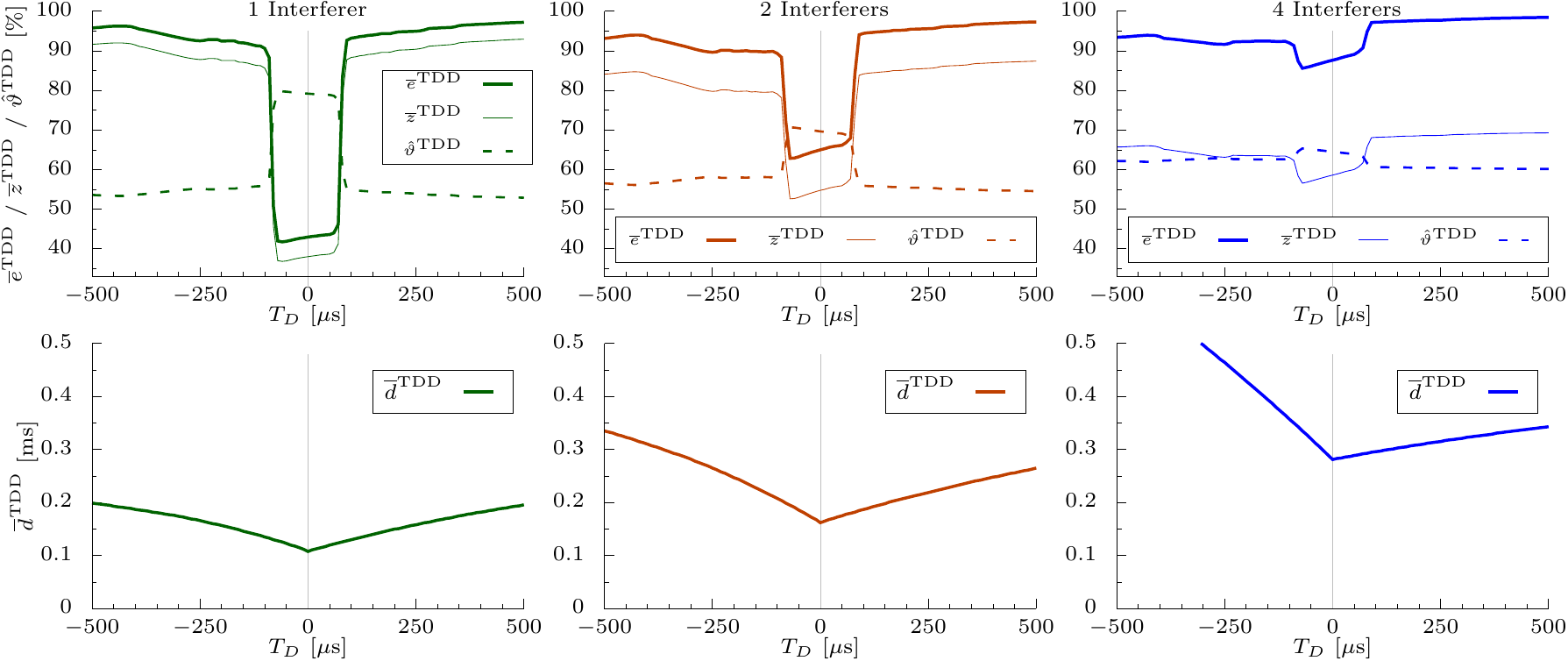}
  \caption{Fractions of early terminated ($\overline{e}^{\mathrm{TDD}}$) and simplex ($\overline{z}^{\mathrm{TDD}}$) transmissions, relative load ($\hat\vartheta^{\mathrm{TDD}}$), and mean latency ($\overline{d}^{\mathrm{TDD}}$), vs. $T_D$ (for TDD).}
  \label{fig:pda}
\end{figure*}

Results for the three runs are reported in Fig.~\ref{fig:pda}.
Plots in the upper part of the figure show some metrics about bandwidth consumption, i.e., the fraction of early terminated packets ($\overline{e}^{\mathrm{TDD}}$), the fraction of simplex packets ($\overline{z}^{\mathrm{TDD}}$), and the relative network load with respect to PoW ($\hat{\vartheta}^{\mathrm{TDD}}$).
Instead, those in the lower part concern the average latency on the redundant link ($\overline{d}^{\mathrm{TDD}}$).

As can be seen, in each run there are two thresholds for $T_D$, located around $\unit[\pm100]{\mu s}$, within which spectrum consumption does not vary appreciably, this meaning that TDD adoption does not bring any improvement over pure RDA.
However, passed those thresholds, behavior improves suddenly, as witnessed by, e.g., the noticeable increase of $\overline{e}^{\mathrm{TDD}}$ values.
This is because transmission requests on the secondary channel are deferred enough by TDD, so that most of them can be prevented by the LRE.
TDD benefits are higher when channels are not heavily loaded, as in the \emph{1\,Int.} case, mostly because saving due to RDA alone is less tangible.
In this respect, TDD adoption improves Wi-Red stability versus network traffic.

In theory, as long as queuing phenomena in Wi-Fi adapters are negligible, parameter $T_D$ has no effect on the amount of lost packets in TDD, which is the same as RDA and PoW.
Instead, it affects transmission latencies, as they are measured from the time the transmission request is issued on the primary channel.
As can be seen in the plots of Fig.~\ref{fig:pda}, mean latency $\overline{d}^{\mathrm{TDD}}$ progressively worsens as $\left|T_D\right|$ increases (in either direction).
This is quite obvious, since TDD intentionally delays one of the two channels.

Plots about $\overline{d}^{\mathrm{TDD}}$ are not symmetric about the y-axis.
Their shape (but not the values) in the different runs is mostly the same when $T_D > 0$.
However, this is no longer true when $T_D < 0$.
This can be explained by remembering that, in the latter case, transmissions are delayed on $A$, and so a higher number of packets arrive first on $B$ when $\left| T_D \right|$ is increased.
As a consequence, communication quality on $B$ matters more.
When interference on that channel is higher, as in the \emph{4\,Int.} case, the related latency is larger.
This implies that, for negative $T_D$ values, the slope of the plots concerning the latency on the redundant link is steeper.

The optimal values for parameter $T_{D}$ are located just above said thresholds, and approximately correspond to the overall duration on air of the initial attempt for $m_i$ on the primary channel (including DATA and ACK frames, SIFS, and $T_\mathrm{LRE}$), plus a safety margin.
With this arrangement, all packet transmissions that find such channel idle and immediately succeed (a non-unusual condition in a well-dimensioned Wi-Fi network, as experimentally shown in \cite{2007-TII-WLAN}) are not replicated on the secondary channel.
This particular choice provides significant benefits concerning spectrum consumption, by keeping the latency increase, on average, to the minimum.

\begin{table}[t]
  \caption{Experimental results for TDD with optimal $T_D$ value.}
  \label{tab:TDD2}
  \centering
  \includegraphics[width=1\columnwidth]{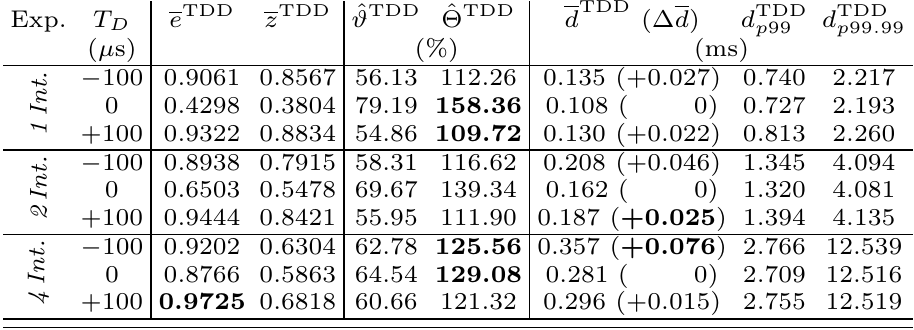}
\end{table}

For example, if $T_D$ is set to either $\unit[-100]{\mu s}$ or $\unit[+100]{\mu s}$ (near to the optimal values for our experimental conditions), the values in Table~\ref{tab:TDD2} are obtained.
With these configurations, $\overline{e}^{\mathrm{TDD}}$ was always higher than $\sim\!90\%$ for all the experiments reported in the plots (and, in the \emph{4\,Int.} case, it exceeded $97\%$ when $T_D = \unit[+100]{\mu s}$).
This means that less than $10\%$ of the packets did not incur in early termination.
At the same time, $\overline{d}^{\mathrm{TDD}}$ never grown more than $\unit[76]{\mu s}$ with respect to pure RDA (see the \emph{4\,Int.} case, $T_D = \unit[-100]{\mu s}$).
In the case adaptive TDD algorithms were implemented, able to dynamically select the best primary channel, $\overline{d}^{\mathrm{TDD}}$ increase would likely remain below $\unit[25]{\mu s}$ (see the \emph{2\,Int.} case, $T_D = \unit[+100]{\mu s}$).

\section{Conclusion}
\label{sec:Conclusion}
Seamless redundancy applied to IEEE 802.11 greatly improves its reliability and makes it suitable for time-sensitive control applications, preserving backward compatibility with Wi-Fi.
When PRP is layered directly above wireless adapters, as in the PoW approach, communication quality improves dramatically, but spectrum consumption becomes twice as much as non-redundant \mbox{Wi-Fi}.
The Wi-Red proposal exploits synergies between the MAC retransmission mechanism and channel redundancy in order to improve behavior further.
In particular, duplication avoidance mechanisms reduce spectrum consumption by preventing unnecessary transmission attempts.

This paper specifically focuses on the ability of such mechanisms to cut down the additional traffic implied by seamless redundancy.
Suitable metrics have been defined to this purpose, which were evaluated experimentally on a real testbed.
In order to be effective, early termination of packet transmissions has to be implemented in the hardware/firmware of Wi-Fi adapters, which is currently out of our reach.
For this reason, an approximate method was devised, where experimental data are acquired on a real testbed that implements PRP on commercial \mbox{Wi-Fi} devices and then processed under the (verified) hypotheses of independence between subsequent transmissions on air and slowly-varying spectrum conditions.

Although in this way we can only obtain a lower bound on the amount of bandwidth saved by duplication avoidance mechanisms (which means that the related results are pessimistic), improvements we measured are nevertheless noteworthy.
Reactive approaches, which cancel transmission of duplicate copies upon reception of the related ACK on at least one channel, provide all the benefits of PoW (and often behave even better in terms of improved reliability and timeliness) but using less bandwidth.
In our experiments, RDA traffic (or, better, its upper bound as given by $\hat{\Theta}^{\mathrm{RDA}}$) was $29.08\%$ to $58.36\%$ higher than what generated, on average, by non-redundant Wi-Fi.
For the proactive TDD approach, additional spectrum consumption with respect to Wi-Fi, as per $\hat{\Theta}^{\mathrm{TDD}}$, lay in the range from $9.72\%$ to $25.56\%$.
In the TDD case, the price to pay is a slight increase in the transmission latency (some tens of $\unit[]{\mu s}$).

In a world where the wireless spectrum is more and more congested, adoption of seamless redundancy to increase \mbox{Wi-Fi} reliability practically mandates the inclusion of duplication avoidance mechanisms.
As future work, we plan to investigate adaptive PDA solutions, aimed at providing optimal performance in spite of variations of the generated traffic pattern and the wireless spectrum conditions, and to analyze networks including multiple RSTAs.

\bibliographystyle{IEEEtran}
\bibliography{TW-Mar-18-0281}

\begin{IEEEbiography}%
[{\includegraphics[width=1in,height=1.25in,clip,keepaspectratio]
{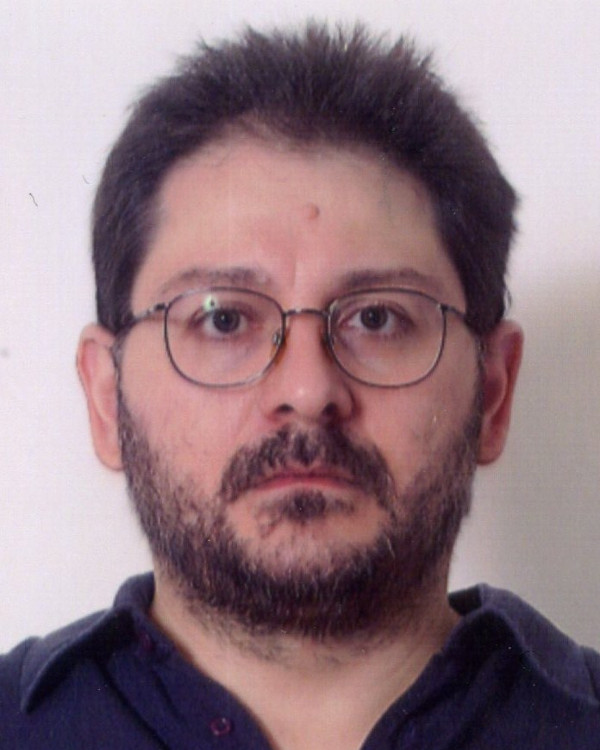}}]{Gianluca Cena} (SM'09) received the Laurea degree in electronic engineering and the Ph.D. degree in information and system engineering from the Politecnico di Torino, Italy, in 1991 and 1996, respectively. Since 2005 he has been a Director of Research with the Institute of Electronics, Computer and Telecommunication Engineering, National Research Council of Italy (CNR--IEIIT), Torino.

His research interests include wired and wireless industrial communication systems, real-time protocols, and automotive networks. In these areas he has coauthored about 130 technical papers, three of which awarded as Best Papers of the 2004, 2010, and 2017 editions of the IEEE Workshop on Factory Communication Systems, and one as 2017 Best Paper for the \textsc{IEEE Transactions on Industrial Informatics}, plus one international patent.

Dr. Cena served as a Program Co-Chairman for the 2006 and 2008 editions of the IEEE International Workshop on Factory Communication Systems, and as a Track Co-Chairman in six editions of the IEEE International Conference on Emerging Technologies and Factory Automation. Since 2009 he has been an Associate Editor of the \textsc{IEEE Transactions on Industrial Informatics}.
\end{IEEEbiography}

\begin{IEEEbiography}%
[{\includegraphics[width=1in,height=1.25in,clip,keepaspectratio]
{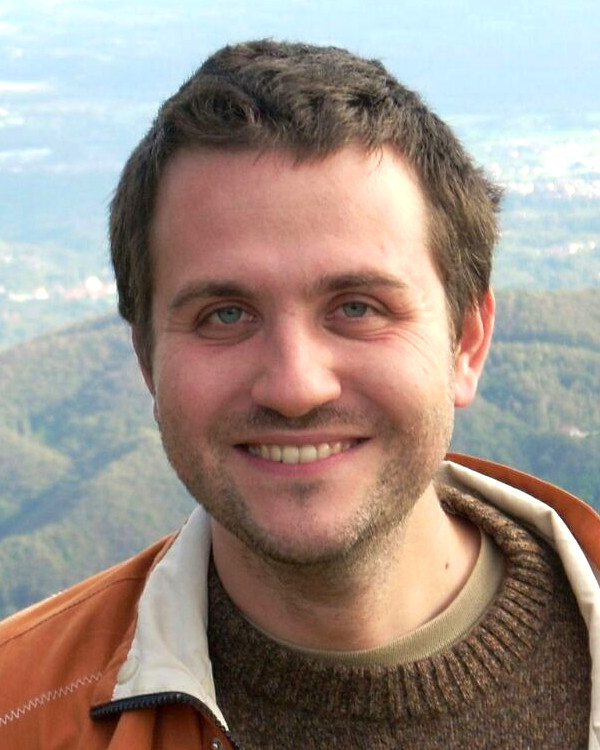}}]{Stefano Scanzio} (S'06-M'12) received the Laurea and Ph.D. degrees in Computer Science from Politecnico di Torino, Torino, Italy, in 2004 and 2008, respectively.
He was with the Department of Computer Engineering, Politecnico di Torino, from 2004 to 2009, where he was involved in research on speech recognition and, in particular, he has been active in classification methods and algorithms. Since 2009, he has been with the National Research Council of Italy (CNR), where he is a tenured Researcher with the Institute of Electronics, Computer and Telecommunication Engineering (IEIIT), Torino.

Dr. Scanzio served as a Work-in-Progress Co-Chairs in the 2018 edition of the IEEE International Workshop on Factory Communication Systems (WFCS 2018). He teaches several courses on Computer Science at Politecnico di Torino. He has authored and co-authored of more than 50 papers in international journals and conferences, in the area of industrial communication systems, real-time networks, wireless networks and clock synchronization protocols. He received the 2017 Best Paper Award for the \textsc{IEEE Transactions on Industrial Informatics}, and the Best Paper Awards for the papers he presented at the 8th and 13th IEEE Workshops on Factory Communication Systems (WFCS 2010 and WFCS 2017).
\end{IEEEbiography}

\begin{IEEEbiography}%
[{\includegraphics[width=1in,height=1.25in,clip,keepaspectratio]
{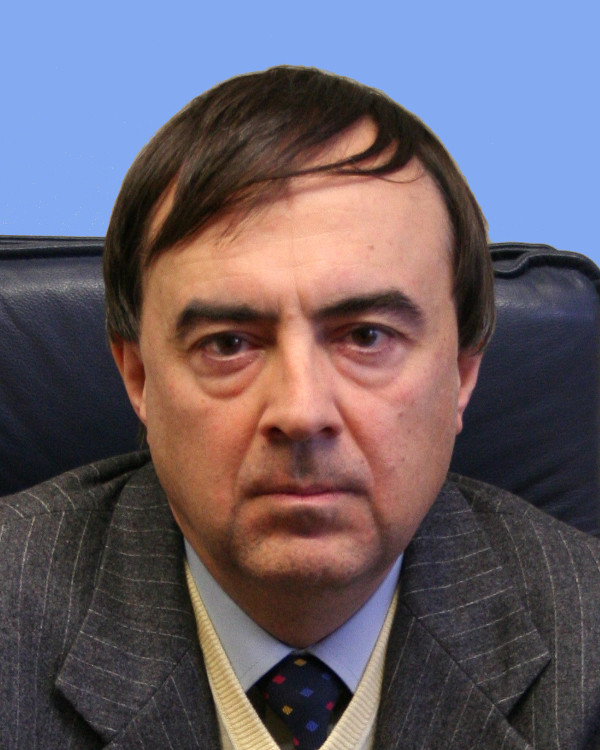}}]{Adriano Valenzano} (SM'09) received the Laurea degree magna cum laude in electronic engineering from Politecnico di Torino, Torino, Italy, in 1980. He is Director of Research with the National Research Council of Italy (CNR). He is currently with the Institute of Electronics, Computer and Telecommunication Engineering (IEIIT), Torino, Italy, where he is responsible for research concerning distributed computer systems, local area networks, and communication protocols. He has coauthored approximately 200 refereed journal and conference papers in the area of computer engineering.

Dr. Valenzano is the recipient of the 2013 IEEE IES and ABB Lifetime Contribution to Factory Automation Award. He was also awarded for the best paper published in the \textsc{IEEE Transactions on Industrial Informatics} during 2016, and received the Best Paper Awards for the papers presented at the 5th, 8th and 13th IEEE Workshops on Factory Communication Systems (WFCS 2004, WFCS 2010 and WFCS 2017).

Adriano Valenzano has served as a technical referee for several international journals and conferences, also taking part in the program committees of international events of primary importance. Since 2007, he has been serving as an Associate Editor for the \textsc{IEEE Transactions on Industrial Informatics}.
\end{IEEEbiography}

\end{document}